\documentclass[preprint,12pt]{elsarticle}
\usepackage[none]{hyphenat}

\usepackage{lineno,hyperref}
\pdfoutput=1
\usepackage{graphics}
\usepackage[english] {babel}
\usepackage{amssymb,amsfonts,amsmath,mathtext,amsthm}
\usepackage{graphicx}
\usepackage{xcolor}
\usepackage{enumitem}
\setlist[enumerate]{itemsep=0mm}
\setlist[itemize]{itemsep=0mm}

\newtheorem{proposition}{Proposition}[section]

\theoremstyle{definition}
\newtheorem{definition}{Definition}[section]

\usepackage{subfig}
\usepackage{floatrow}
\usepackage{nicefrac}
\usepackage{float}
\floatstyle{plaintop}
\restylefloat{table}
\usepackage{nicefrac}
\usepackage{undertilde}

\usepackage{natbib}
\setcitestyle{authoryear,open={(},close={)}}

\DeclareMathOperator{\Ex}{\mathbb{E}}
\newcommand{\Var}{{\mathsf{Var}}\,}
\newcommand{\R}{{\mathbb R}}

\renewcommand{\S}{{\mathbb S}}

\newcommand{\constant}{{\rm{const}}}

\newcommand{\TT}{\textsf{T}}
\DeclareMathOperator*{\argmin}{argmin}

\renewcommand{\d}{{\rm d}}

\newcommand{\eg}{e.g.,\ }

\newcommand{\cf}{{\em cf.\ }}
\newcommand{\vs}{vs.\ }
\newcommand{\ie}{i.e., }

\newcommand{\given}{{\mbox{$\,|\,$}}}

\newcommand{\e}{\,\mathrm{e}}
\renewcommand{\i}{\mathsf{i}}

\modulolinenumbers[5]

\journal{Spatial Statistics}

\begin{document}

\begin{frontmatter}

\title{Regularization of the ensemble Kalman filter using a non-parametric, non-stationary spatial model\footnote{
This is the authors' final version of the paper accepted for publication in 
    Spatial Statistics.   DOI: \url{https://doi.org/10.1016/j.spasta.2024.100870} \\
\textcopyright 2024. This manuscript version is made available under the CC-BY-NC-ND 4.0 license \url{https://creativecommons.org/licenses/by-nc-nd/4.0/}
}
}



\author[main]{Michael Tsyrulnikov\corref{mycorrespondingauthor}}
\ead{michael.tsyrulnikov@gmail.com}

\affiliation[main]{organization={HydroMeteorological Research Center of Russia},
            country={Russian Federation}}

\cortext[mycorrespondingauthor]{Corresponding author}

\author[main,scnd]{Arseniy Sotskiy}
\affiliation[scnd]{organization={National Research University Higher School of Economics},
            country={Russian Federation}}

\begin{abstract}

The sample covariance matrix of a random vector is a 
good estimate of the true covariance matrix if the sample size is much larger than the length of the  vector. 
In high-dimensional problems, this condition is never met. As a result, in high dimensions 
the Ensemble Kalman Filter's (EnKF) ensemble does not contain enough 
information to specify the prior covariance matrix accurately. 
This necessitates the need for regularization of the analysis (observation update) problem.
We propose a regularization technique based on a new spatial model.
The model is a constrained version of the general Gaussian process convolution model.
The constraints include local stationarity and smoothness of local spectra.
We regularize EnKF by postulating that its prior covariances obey the spatial model. 
Placing a hyperprior distribution on the model parameters and using the likelihood of the prior ensemble data
allows for an optimized use of both the ensemble and the hyperprior.
A linear version of the respective estimator is shown to be consistent. A more accurate
nonlinear neural-Bayes implementation of the estimator is developed.
In simulation experiments, the new technique led to substantially better EnKF performance than several existing techniques.

\end{abstract}


\end{frontmatter}


\section {Introduction}
\label{sec_intro}

Modern  data assimilation in geosciences and related areas 
relies on the ensemble approach,
in which the probability distribution of the unknown system state is represented by
a finite sample (ensemble) of pseudo-random realizations (ensemble members).
This allows for realistic, spatially varying  prior covariances in the analysis
(the observation update step of the data assimilation cycle).
In practical applications, the  most widely-used approach is the Ensemble Kalman Filter (EnKF)
and its numerous modifications and extensions\footnote{
See  section \ref{sec_seq_flt} for a brief introduction 
and \ref{App_lin_seq_flt} for a more technical description
of sequential ensemble filtering in the linear setting (which is sufficient
for our purposes in this study).
}.
The principal problem of the ensemble approach is  that maintaining a large ensemble in 
the course of data assimilation (sequential filtering)  is
computationally expensive. 
For real-world systems modeled with high-resolution models with billions of degrees of freedom,
this means that only relatively small ensembles are affordable.
As a result, such ensembles can  provide the analysis  with
only scarce information on the  prior distribution. The sample covariance matrix 
is a poor estimate of the true covariance matrix if the sample size
is much lower than the dimensionality of the state space.
This requires a kind of {\em regularization}, that is, the introduction of
additional information on the prior covariances.

\subsection {Related work}
\label{sec_intro_literature}

The following EnKF regularization techniques have been proposed so far.
\begin{enumerate}
\item
\label{list_lcz}
The most popular approach is covariance
localization (tapering), \citep[e.g.][]{Furrer,Houtekamer2016},
which reduces spurious long-distance correlations through element-wise multiplication
of the sample covariance matrix by an ad-hoc positive definite localization matrix.
This technique efficiently removes a lot of noise in the sample covariance matrix.
However, it cannot cope with the noise at small distances (where the noise is the strongest). 
And it inevitably leads to an 
underestimation of the  assumed length scale of the forecast-error field and thus to imbalances 
between different fields \citep{LorencEnKF}.


\item
\label{list_hybr}
Blending prior sample covariances
with static (time-mean) covariances is  widely used in meteorological ensemble-variational schemes
\citep{Buehner2013,Lorenc}. In statistical literature, similar techniques are known
as shrinkage estimators \citep{Ledoit}.
Sample covariances are noisy but they do contain the useful flow-dependent `signal'.
Static covariances are noise-free but can be irrelevant under some flow conditions and 
observation coverage. A convex combination of the two covariance matrices proved to be useful
(see the above references) but
it is not selective: the noise in the sample covariances is reduced
to the same extent as the flow-dependent non-stationary signal.

\item
\label{list_spat_ave_covs}
Another  approach is spatial averaging of sample covariances (that is, blending
with neighboring in space covariances), \eg \citet{Berre2010}.
The technique reduces noise in sample covariances
due to an increase in the effective ensemble size but it does so
at the expense of somewhat distorting (smoothing) the covariances.
Optimal linear filtering of sample covariances \citep{Menetrier} further develops
this idea.

\item
\label{list_intro_recent}
A similar technique is temporal averaging of prior sample covariances
(\ie blending with recent past covariances).
\citet[][]{Berre2015} and \citet[][]{Bonavita} used prior ensembles from several previous 
assimilation cycles to increase the ensemble size.
\citet[][]{Lorenc2017} used time-shifted ensemble perturbations
for the same purpose.
\citet{HBEF}  arrived at this technique by  assuming
that the true covariance matrix is an
unknown random matrix  and introducing a secondary filter in which the covariances are updated.

\item
\citet{Ueno} proposed to regularize the sample covariance matrix by imposing
a sparse structure in the inverse covariance (precision) matrix.
A similar approach was taken  by \citet{Hou}.
\citet{Tjelmeland2023} imposed a Markov random field structure to obtain a sparse 
prior precision matrix. 
\citet{Katzfuss2019} postulated sparsity in inverse Cholesky factors of the prior
covariance matrix.

\item
\label{list_prm_cov}
One more option is to adopt a parametric model for the prior covariances and estimate
parameters of the model from the forecast ensemble. \citet{Skauvold} 
used  maximum likelihood estimates of a few model parameters
in stationary and non-stationary settings.

\item
\label{list_wvl_mdl}
Finally, imposing a structure in wavelet space
can be used to regularize the analysis.
Postulating the simplest structure, the diagonal wavelet covariance matrix,
 proved to be effective in data assimilation applications
\citep{Fisher,Pannekoucke2007}, see also \citet{Mandel}.
This indicates that the regularizing effect
of that approach can outweigh methodological errors related to 
the neglect of off-diagonal covariances in wavelet space.

\end{enumerate}

\subsection {Motivation and contributions}
\label{sec_intro_motiv}

In this paper we consider systems whose state vector represents a {\em spatial field}, \eg atmosphere,
ocean, and the like.
Taking advantage of this interpretation of the state vector,
we propose to regularize the analysis problem by imposing a {\em spatial model}. In this regard,
our approach is similar to the two latter techniques  (items \ref{list_prm_cov} and \ref{list_wvl_mdl} above).
In contrast to parametric covariance models, we aim at a non-parametric formulation
to obtain a more flexible model.
By non-parametric, we mean defined by parameters whose number is not fixed (as with a parametric model)
but grows with the increasing spatial resolution of the model.
Compared to the wavelet-diagonal models, which provide regularizing information
by imposing zero cross-correlations  between wavelet coefficients, 
our model does so by imposing several constraints
on the general process convolution model. The two principal constraints are local stationarity
(under which a spatially non-stationary process can be approximated by a stationary process locally in space)
and smoothness of what we call local spectra.

We assume that the prior distribution of the state at each filtering time step
is multivariate Gaussian with the mean equal to the ensemble mean 
(or, for a linear forecast model, to the deterministic forecast,
see Remark in \ref{App_anls_ens}). The prior covariances are postulated to be governed by
the spatial model introduced in this paper. 
We place a hyperprior distribution on the model parameters and
use the likelihood of the ensemble data to online estimate the spatial model. 
The estimated model is then employed to compute 
a square root of the prior covariance matrix, which is finally used in the analysis.

The reason why we propose to rely on a Bayesian estimate of a spatial model in the EnKF analysis 
is that a more principled 
(than the above ad-hoc  techniques) way of covariance regularization
can lead to a better estimate of the true prior covariance matrix and thus better performance of the filter.
It is also worth noting that the introduction of the hyperprior distribution implies an {\em enhancement} of EnKF that
goes beyond blending of sample covariances with time-mean covariances (mentioned in item \ref{list_hybr} above).
 
Our key contributions are:
\begin{itemize}
\item
A  constrained non-parametric, non-stationary process convolution model on the sphere.
The concept of local stationarity of the resulting spatial process defined by
conditioning on the random convolution kernel.
\item
An estimator of the spatial model suitable for high-dimensional applications.
A proof that a linear version of the estimator is consistent.
A nonlinear, neural Bayes realization of the estimator. 
\item
Implementation of the model and its estimator in EnKF, leading to a new Locally Stationary Ensemble Filter (LSEF).
\item
Numerical experiments with simulated data with the LSEF analysis on the sphere and the LSEF on the circle.
\end{itemize}

The rest of the paper is organized as follows.
Section \ref{sec_seq_flt} gives a brief overview of sequential ensemble filtering.
The theoretical part of the paper consists of two big sections, \ref{sec_conv} and \ref{sec_LSM_estm},
supported by several Appendices.
In section \ref{sec_conv} we introduce  a  non-parametric locally stationary Gaussian process convolution model
on the sphere (and on the circle). 
In section \ref{sec_LSM_estm} we propose an estimator of the model from
an ensemble of random field realizations.
Section \ref{sec_LSEF_anls} describes
how the spatial model is used to regularize the analysis step of the EnKF.
The experimental part of the paper is presented in sections \ref{sec_expm_anls}and  \ref{sec_expm_LSEF}.

Reproducible Python and R code for numerical experiments presented in this paper 
can be found at \url{https://github.com/Arseniy100/LSEF}.

\section{Background on sequential ensemble filtering}
\label{sec_seq_flt}

The goal  of  sequential filtering is the estimation of the current state 
of an evolving {\em system}, having 
(i) a forecast model capable of predicting the future state of the system given its current state, 
(ii) a set of observations  on the system, and 
(iii) an observation model that relates the observations to the system state (the truth).

In the time-discrete filtering, observations are available at discrete time instants.
The estimates of the (ground) truth are sought at the same time instants.
A sequential filter consists of 
alternating short-term forecast (time update) and analysis (observation update) steps.
In the time-discrete {\em linear} filtering, which is sufficient to test the 
covariance regularization technique we propose in this research,
the evolution of the truth ${\bf x}_{t}$ between consecutive time instants $t$
is assumed to be governed by the stochastic equation
\begin {equation}
\label{evolu}
{\bf x}_{t} = {\bf F}_t {\bf x}_{t-1} + \boldsymbol\eta_t,
\end {equation}
where ${\bf F}_t$ is the known forecast-model operator (a matrix) and $\boldsymbol\eta_t$ is the unknown random model error.
The assimilated observations are assumed to be related to the truth as
\begin {equation}
\label{obs} 
{\bf x}^{\rm obs}_t = {\bf H}_t {\bf x}_t + \boldsymbol\varepsilon_t,
\end {equation}
where ${\bf H}_t$ is the known observation-model operator and
$\boldsymbol\varepsilon_t$ is the unknown random observation error.

If the errors $\boldsymbol\eta_t$ and $\boldsymbol\varepsilon_t$ are all mutually uncorrelated and have known
first and second moments, then the mean-square optimal estimate of 
${\bf x}_{t}$ given ${\bf x}^{\rm obs}_t, {\bf x}^{\rm obs}_{t-1}, \dots$
is given by the Kalman filter \citep[e.g.][]{Jazwinski}.
In high dimensions, however, the Kalman filter is prohibitively expensive.
Its Monte-Carlo version known as the Ensemble Kalman filter (EnKF) is an affordable approximate filter, 
\citep[e.g.][]{Asch}.

EnKF propagates in time a sample (ensemble) of `possible truths' consistent with observations. 
A {\em forecast (prior) ensemble} is computed in 
EnKF at every assimilation cycle $t$ before assimilation of current observations, 
and an {\em analysis (posterior) ensemble} is computed after current observations
are assimilated.
The key assumption in EnKF is that the prior ensemble is a sample from the true prior distribution.
EnKF computes a {\em regularized sample covariance matrix} $\widehat{\bf B}_t$
from the prior ensemble and substitutes it
for the {\em true prior covariance matrix} ${\bf B}_t$ in the Kalman-filter analysis equation.
It is the prior sample covariance matrix that we aim to regularize in this research using a spatial model ---
for systems whose state is a spatial field.

Having a deterministic forecast (the mean of the prior ensemble or the so-called control forecast),
we deal in what follows  with the  the {\em forecast-error} random field, that is,
the deviation of the forecast from the truth.
Online information on its current (\ie at the time instant $t$) probability distribution is given by  
 {\em ensemble perturbations}, the differences of the prior ensemble members 
and the deterministic forecast.

For more details on approximate sequential filtering, see  \ref{App_lin_seq_flt}.

\section {Spatial model}
\label{sec_conv}

We employ a non-parametric spatial model for the 
forecast-error random field to regularize the data assimilation problem.
Following  \citet{Higdon}, we rely on the Gaussian {\em process convolution} model.
In contrast to most applications, which postulate
a parametric model for the spatial convolution kernel \citep[e.g.][]{Sanso2009,Bhat,Katzfuss,Li}, we
choose a {\em non-parametric} approach to allow for variable shapes of spatial covariances.
Non-parametric convolution kernels were considered by  \citet{Tobar2015}
and  \citet{Bruinsma2022} in the  stationary time series context. 
Here we propose a non-parametric non-stationary spatial convolution model on compact manifolds such as 
the sphere and the circle.

The default domain is the two-dimensional unit sphere $\S^2$ 
(the domain of interest in geosciences). We also formulate and test the model 
on the unit circle $\S^1$ (used as a lower-dimensional counterpart of the sphere).
We refer to $\S^2$ and $\S^1$ just as the sphere and the  circle in the sequel.
We denote points on the sphere as $x,y,\dots$ and area elements simply by $\d x, \d y,\dots$
On the sphere, we use the terms stationarity and isotropy interchangeably.
The terms field and process are also used interchangeably.
We confine ourselves to band-limited processes because in the context of sequential filtering, 
the spatial field to be estimated 
inevitably has limited spatial resolution.
We also assume that probability distributions we deal with in this study 
have densities with respect to the Lebesgue measure.


The rest of this section is structured as follows.
We start by defining our Gaussian process convolution model, including the introduction of  so-called local spectra.
Then, we realize that  to be identifiable from the prior ensemble,  the model needs to be constrained.
We propose the following four constraints:
(i) local isotropy of the convolution kernel,
(ii) positive definiteness of the non-stationary convolution kernel as a function of distance,
(iii) local stationarity,
(iv)  smoothness of the local spectra.

\subsection {General process convolution model}
\label{sec_conv_gen}

Let $\xi(x)$ be a general real-valued space-continuous zero-mean linear Gaussian process:
\begin {equation}
\label{xiw}
\xi(x) = \int_{D} w(x, y) \,\alpha(y) \,\d y \equiv 
          \int_{D} w(x, y) \,Z(\d y).
\end {equation}
Here 
$D$ is the domain of interest,
$\alpha$ is the white Gaussian noise,
$Z$ is the  Gaussian orthogonal stochastic measure 
such that $\Ex Z(\d x)=0$, $\Ex Z^2(\d x)=|\d x|$,   
$\Ex Z(\d x) Z(\d y)=0$ whenever $\d x \cap \d y =\emptyset$,   
$\Ex$ is expectation, and
$w(x, y)$ is a non-random real function called the convolution kernel.

The (non-stationary in general)
covariance function of the process $\xi(x)$ defined by Eq.~(\ref{xiw}) is
$B(x,x')=\Ex\xi(x)\xi(x') = \int_{D} w(x, y) w(x', y) \,\d y$.

\subsection {Constraining the model: strategy}

In the formulation Eq.~(\ref{xiw}), the model is not identifiable, that is, 
the function $w(x, y)$ is not unique given the process covariance function  $B(x,x')$. 
Indeed, consider an isotropic convolution kernel $\psi(\rho(y,y'))$,
where $\rho(y,y')$ stands for the great-circle distance between the points $y$ and $y'$.
Let the Fourier-Legendre transform of $\psi(\rho)$ be equal to one in modulus 
(see  Eq.~(\ref{u_spe}) for the definition of the Fourier-Legendre transform we use). 
Then, it is straightforward to see
that the covariance function remains unchanged if $w(x, y)$ is convolved
with $\psi$\footnote{
By the convolution of $w(\cdot, \cdot)$ with $\psi(\rho(\cdot,\cdot))$ we mean 
$w_*(x,y) =\int w(x, z) \, \psi(\rho(z,y)) \,\d z$.
}.
(The convolution with such a kernel is analogous to the multiplication by an orthogonal matrix
in the space-discrete case.)

We intend to introduce  constraints on the convolution kernel to make the model identifiable and
reduce the effective number of degrees of freedom in the model,
thus facilitating its estimation from an ensemble of random field realizations.

\subsection {Locally isotropic kernel}
\label{sec_conv_loc_isotr}

With a fixed $x$, we require that $w(x, y)$ depends on $y$ only through the great-circle distance  $\rho(x, y)$
between the two points. This is the {\bf first constraint} we impose on the convolution kernel.
It implies that there is a real-valued function  $u(x, \rho(x,y))$ such that
\begin {equation}
\label{wv}
w(x, y)  = u(x, \rho(x,y)).
\end {equation}
Substituting  Eq.~(\ref{wv}) into Eq.~(\ref{xiw}) yields
\begin {equation}
\label{osc2}
\xi(x) = \int_D u(x, \rho(x,y)) \,\alpha(y) \,\d y. 
\end {equation}
In the remainder of this subsection, we develop spectral representations of the process 
defined by Eq.~(\ref{osc2}) and its covariance function.
We perform the spectral (Fourier-Legendre)
expansion of $u(x, \rho)$  with $x$ being fixed:
\begin {equation}
\label{u_spe}
u(x, \rho) =   \sum^{\ell_{\rm max}}_{\ell=0} \frac{2\ell +1}{4\pi}  \sigma_\ell(x) P_\ell(\cos \rho),
\end {equation}
where $\ell_{\rm max}$ is the bandwidth,
$\sigma_\ell(x)$ are real-valued functions (we call them {\em spectral functions})
and $P_\ell$ are the Legendre polynomials.
We substitute $\rho=\rho(x,y)$ into Eq.~(\ref{u_spe}) and apply
the addition theorem of spherical harmonics, 
\begin {equation}
\label{addThm}
\sum_{m=-\ell}^\ell Y_{\ell m}(x) Y_{\ell m}^*(y) = \frac{2\ell +1}{4\pi} P_\ell(\cos \rho(x,y)),
\end {equation}
where $Y_{\ell m}$ is the spherical harmonic, $\ell$ its degree (sometimes called the total wavenumber),
$m$ its order (the meridional wavenumber), and the asterisk denotes complex conjugation.

As a result, Eq.~(\ref{u_spe}) can be rewritten as
\begin {equation}
\label{u_spe2}
u(x, \rho(x,y)) =  \sum^{\ell_{\rm max}}_{\ell=0} \sigma_\ell(x) \sum^\ell_{m=-\ell} 
                           Y_{\ell m} (x) \, Y_{\ell m}^* (y).
\end {equation}
Then, we make use of the spectral expansion of the white noise, 
\begin {equation}
\label{alpha}
\alpha(y) = \sum^{\ell_{\rm max}}_{\ell=0} \sum^\ell_{m=-\ell}  \widetilde\alpha_{\ell m}  \, Y_{\ell m} (y),
\end {equation}
where $\widetilde\alpha_{\ell m}$ are mutually uncorrelated
complex-valued random variables with 
$\Ex\widetilde\alpha_{\ell m}=0$ and $\Var\widetilde\alpha_{\ell m}=1$.
More specifically, $\widetilde\alpha_{l0}$ are real-valued and all the other $\widetilde\alpha_{\ell m}$
 are complex circularly symmetric
random variables \citep[e.g.][section 3.7]{Gallager} such that 
$\widetilde\alpha_{l,-m} = \widetilde\alpha_{\ell m}^*$.

Substituting Eqs.~(\ref{u_spe2}) and (\ref{alpha}) into (\ref{osc2}) and
utilizing orthonormality of spherical harmonics, we  obtain the basic spectral
representation of the random process that satisfies Eq.~(\ref{osc2}):
\begin {equation}
\label{LSM}
\xi (x) = \sum^{\ell_{\rm max}}_{\ell=0} \sum^\ell_{m=-\ell}  \sigma_\ell(x)  
                          \,\widetilde\alpha_{\ell m} \, Y_{\ell m} (x).
\end {equation}
The spatial covariances $B(x,x') = \Ex \xi (x) \xi (x')$ can readily be  obtained from Eq.~(\ref{LSM}) by taking into account
that all $\widetilde\alpha_{\ell m}$ are mutually uncorrelated and
again applying  the addition theorem of spherical harmonics.
Equivalently, we may use Eq.~(\ref{osc2}) for this purpose so that
\begin {equation}
\label{BxxS2}
B(x,x')  = \int_D u(x, \rho(x,y))  \, u(x', \rho(x',y))  \,\d y =
 \sum^{\ell_{\rm max}}_{\ell=0}  \frac{2\ell +1}{4\pi}   
                          \,\sigma_\ell(x) \, \sigma_\ell(x')\, P_\ell(\cos \rho(x,x')).
\end {equation}
If $u(x, \rho) = u(\rho)$ or, equivalently, if all spectral functions $\sigma_\ell(x)$ do not depend on $x$,
then $\xi(x)$ becomes stationary, in which case the process variance 
$\Var \xi (x)$ can be obtained by summing 
$\frac{2\ell +1}{4\pi} \sigma_\ell^2$ known as the variance spectrum  (energy per degree $\ell$).
Besides, $f_\ell = \sigma_\ell^2$ is often called the 
{\em modal} spectrum  (energy per `mode', \ie per pair $(\ell,m)$).
In the non-stationary case, in which $\sigma_\ell$ depends on $x$, we call 
$f_\ell(x) = \sigma_\ell^2(x)$ the {\em local spectrum}.
Equation (\ref{BxxS2}) shows that the non-stationary process variance 
can be computed in the same way as the stationary one: 
$\Var \xi (x) = \sum_\ell \frac{2\ell +1}{4\pi} \sigma_\ell^2(x) \equiv \sum_\ell \frac{2\ell +1}{4\pi} f_\ell(x)$.

Non-negative definiteness (\ie validity) of the model's non-stationary covariance function 
$B(x,x')$ is established in \ref{App_pos_def}.

The formulation of the spatial model Eq.~(\ref{LSM}) on the circle is outlined in  \ref{App_LSM_S1}.

\subsection {Positive definite convolution kernel}
\label{sec_pos_def_kernel}

In  \ref{App_kernel_lcz}, we consider the process $\xi_{\rm statio}$ defined 
by the process convolution model with the {\em stationary} kernel $u(\rho)$, see Eq.~(\ref{xi_statio}).
As discussed above, the kernel is not unique given the covariance function of the process. 
In an attempt to select a unique  kernel, we impose a computationally motivated 
{\em spatial localization} requirement, seeking $u(\rho)$  that has the smallest spatial scale.
We show that there are three solutions to this optimization problem,  all of them having the same
shape of the kernel.
One of these three equivalent solutions is {\em positive definite}, which we select.
So, if a stationary process with a given covariance function is
defined by the convolution $\xi=u*\alpha$
(where  $\alpha$ is the white noise), then the most spatially localized kernel $u$
can be taken to be a  positive definite function of distance.

Motivated by this result, we postulate that in the non-stationary case, for any $x$, the 
kernel $u(x,\rho)$ is a {\em positive definite function} of  distance $\rho$.
As a consequence, $\sigma_\ell(x) > 0$  both in the spherical and 
in the circular case.
This constitutes our {\bf  second constraint} imposed on the general 
process convolution model.

\subsection {Uniqueness of the convolution kernel}
\label{sec_uniq_ker}

In   \ref{App_identif} we prove that under the local isotropy 
constraint (section \ref{sec_conv_loc_isotr}),  the positive-definite-kernel constraint
(section \ref{sec_pos_def_kernel}), and a technical assumption,   
the kernel $u(x,\rho)$ is uniquely determined 
by the model's non-stationary covariance function $B(x,x')$.

Aiming to further reduce the effective number of degrees of freedom, 
we introduce two additional regularizing constraints (presented in the following two subsections). 
Both are smoothness constraints, one imposed in physical space and the other in spectral space.

\subsection {Local stationarity}
\label{sec_LSM_locsta}

The notion of local stationarity has been defined differently by different authors
(most often for  processes on the real line).
The general idea is that a locally stationary process can be approximated by a stationary process
locally, \ie in the vicinity of any point in time \citep{Mallat}.

Starting from \citet{Dahlhaus1997}, the common approach to studying local stationarity is 
the use of the `infill' asymptotics, 
in which the non-stationarity of  the process $\xi(t)$ is determined by a function, $a(\tau)$, defined in
rescaled time $\tau \in[0,1]$.
Then, $\xi(t)$ is considered on $t =  T\tau \in [0,T]$ as $T \to \infty$.
The prototypical example is the autoregressive process 
$\xi_{t+1} =   a(t/T) \,\xi_t +  \eta_t$,
where $t=0,1,\dots,T$ and $\eta_t \sim N(0,1)$.
The mean time-scale of this process does not change as $T \to \infty$, whereas  
the non-stationarity pattern is infinitely stretched.
As a result, for any point $\tau_0$ in rescaled time, $\xi_t$ can be approximated by a stationary process 
in the vicinity of the point $t=T \tau_0$, with the approximation error vanishing
as $T \to \infty$.

On compact manifolds like the circle or the  sphere, this rescaling
cannot be used because of their compactness. 
Therefore, we propose a  different approach to local stationarity with our process convolution model.

Recall that the non-stationary random process in question is defined by Eq.~(\ref{LSM}), where
the non-stationarity comes from the dependencies of $\sigma_\ell$ on the location $x$.
We assume that these are caused by $\sigma_\ell(x)$ {\em being random processes themselves}.
More specifically, we postulate that  $\sigma_\ell(x)$
are, jointly, a vector-valued mean-square differentiable {\em stationary random process}.
Then, we obtain spatial non-stationarity  by  considering
the conditional distribution of the process $\xi$ given all $\sigma_\ell(\cdot)$.
We denote this conditioning by $\xi  \,|\, \sigma$.
In other words, we model spatial non-stationarity by assuming that 
$\boldsymbol\sigma(x) = \{\sigma_\ell(x)\}_{\ell=0}^{\ell_{\rm max}}$ is a random process and examine 
$\xi(x) = \sum  \sigma_\ell(x) \,\widetilde\alpha_{\ell m} \, Y_{\ell m} (x)$ with 
a random realization of $\boldsymbol\sigma(x)$ being fixed.
Equivalently, conditioning on the convolution kernel  $u(x,\rho)$ 
(viewed as a Hilbert space-valued stationary random process of $x$) can be considered.

We  define the length scale $\Lambda_\ell$ of the stationary 
process $\sigma_\ell(x)$ as the micro-scale from the equation 
\begin {equation}
\label{Lsigma}
\Ex \left| \frac{\partial \sigma_\ell}{\partial x} \right|^2  = \frac{\Var\sigma_\ell}{\Lambda_\ell^2}
\end {equation}
(where $\partial / \partial x$ is the spatial gradient operator)
and the overall {\em non-stationarity length} scale  $\Lambda$ from 
\begin {equation}
\label{Lsigma2}
\sum_{\ell=-\ell_{\rm max}}^{\ell_{\rm max}} \frac{2\ell +1}{4\pi}
  \frac{\Var\sigma_\ell}{\Lambda_\ell^2} = \frac{\Ex \xi^2}{\Lambda^2}.
\end {equation}
It is straightforward to prove that the process $\xi(x)$ is indeed locally stationary in the sense that for any point $x_0$,
it can be approximated by a stationary process $\zeta(x; x_0)$, with the
mean-square approximation error vanishing as $\Lambda \to\infty$.
To do so, we define 
$\zeta(x; x_0) = \sum_{\ell}  \sigma_\ell(x_0) \sum_m \,\widetilde\alpha_{\ell m} \, Y_{\ell m} (x)$,
that is, with the argument of $\sigma_\ell$ fixed at $x_0$.
Then, the mean approximation error variance  is
\begin {multline}
\label{VD}
\Ex\Ex [(\xi(x) - \zeta(x; x_0))^2 \,|\, \sigma] = 
  \sum_{\ell=-\ell_{\rm max}}^{\ell_{\rm max}} \frac{2\ell +1}{4\pi} \Ex\left[\sigma_\ell(x) - \sigma_\ell(x_0) \right]^2 \le\\
   \rho^2(x,x_0) \sum_{\ell=-\ell_{\rm max}}^{\ell_{\rm max}} \frac{2\ell +1}{4\pi}
    \Ex\left| \frac{\partial \sigma_\ell}{\partial x} \right|^2 =
  \frac {\rho^2(x,x_0) }{\Lambda^2}\Ex \xi^2. 
\end {multline}
Here the last equality is due to Eqs.~(\ref{Lsigma}) and (\ref{Lsigma2}).

Theoretically, Eq.~(\ref{VD}) shows that in the locally stationary limit, 
as $\Lambda\to\infty$ and $\Ex \xi^2=\constant$, the
approximation error vanishes globally, that is,  for all points $x$ on the sphere.
Practically, we wish a  locally stationary process to have correlations that,
approximately, depend only on the distance $\rho$
for distances {\em within a process length scale}, $L$ (whose definition is application dependent). 
Then, Eq.~(\ref{VD}) implies that in applications, we may assume local stationarity if $L \ll \Lambda$.


{\em Local stationarity} is the {\bf  third constraint} we introduce.
Note that the  slow variation of the kernel $u(x,\rho)$  with location $x$ 
(compared with its variation as a function of  distance $\rho$), actually, justifies the term 
`local spectrum' (section \ref{sec_conv_loc_isotr}) introduced first in the time series context by \citet{Priestley}, who called it
evolutionary spectrum.

\subsection {Smoothness of local spectra}
\label{sec_smoo_spec}

Studies of real-world spatio-temporal processes 
showed that spatial (and temporal) spectra in turbulent flows
are  quite smooth, exhibiting typically a power-law behavior at large wavenumbers
\citep[e.g.][]{GageNastrom,Trenberth}.
For this reason and to regularize the spatial model
by further reducing its effective number of degrees of freedom, we postulate that 
the spectral functions $\sigma_\ell(x)$ are {\em smooth
functions of the wavenumber} $\ell$ --- this is our {\bf fourth  constraint}.
As for a measure of smoothness of the local spectrum, in section \ref{sec_estm_consi}
we propose one definition aimed to facilitate the proof  of consistency of the estimator, 
and in section \ref{sec_nrBayes} we impose another, more practically oriented smoothness constraint
by placing a prior distribution over the functions $\sigma_\ell(x)$.

Note that if  $\sigma_\ell(x)$ is smooth as a function of $\ell$, then its inverse spectral transform,
that is,
the convolution kernel $u(x,\rho)$, rapidly decreases with distance $\rho$.
The rapidly decreasing convolution kernel prohibits spurious 
(and damaging in the analysis) long-distance correlations, which are present in sample 
(ensemble) covariances.

\subsection {Summary of constraints}
\label{sec_constr}

\begin{enumerate}[label={(\arabic*)}]
\item
\label{list_constr_loc_isotr}
The convolution kernel has the locally isotropic form $w(x,y)=u(x,\rho(x,y))$.

\item
\label{list_constr_pos_def}
The convolution kernel $u(x,\rho)$ is a positive definite function of 
 distance $\rho$.

\item
\label{list_constr_smoo_x}
The process $\xi(x)$  is locally stationary.

\item
\label{list_constr_smoo_l}
The spectral functions   $\sigma_\ell(x)$
are smooth functions of the wavenumber $\ell$.

\end{enumerate}

Below we refer to the above spatial model that obeys these four constraints as the Locally Stationary Convolution Model.

\subsection {Spatial discretization}
\label{sec_dscr}

When used in the analysis to specify the prior covariance matrix (see section \ref{sec_LSEF_anls}),
the model needs to be discretized, that is, defined on a spatial grid.
This is done by  approximating the convolution integral in Eq.~(\ref{osc2}) as a sum.
In this proof-of-concept study, we restricted the band-limited field $\xi(\cdot)$ to the
regular latitude-longitude grid described in section \ref{sec_expm_anls_setup}.
In a real-world high-dimensional application, the key point will be to design a numerical scheme in which, 
for each grid point $x$,  the sum runs over a relatively small number of neighboring grid points.
This will make the resulting matrix operator (denoted in section \ref{sec_LSEF_anls} as $\widehat{\bf W}$) {\em sparse}.
Developing such an algorithm will be done in a follow-up study using a multi-scale/multigrid approach.
A non-ensemble version of the multi-scale technique (with static spatial covariances) 
has been successfully tested in an operational meteorological
data assimilation system \citep{A3F}. 
In that application, we employed a non-regular, quasi-uniform grid on the sphere
to maintain sparsity of the $\widehat{\bf W}$ matrix in the polar regions, 
where the latitude-longitude grid is dense.

\section {Estimation of the spatial model from the ensemble}
\label{sec_LSM_estm}

In this section, we propose an estimator of the Locally Stationary Convolution Model from 
an  ensemble (sample)  of $K$ independent identically distributed random field realizations
(ensemble perturbations),
$\{\xi^{(k)} (x)\}_{k=1}^{K}$.
The goal is to estimate the convolution kernel $u(x,\rho(x,\cdot))$, where 
$x$ runs over points of the spatial grid.
In this study, we confine ourselves to {\em point estimates} of $u(x,\rho(x,y))$.
We recall that for any  $x$, the convolution kernel $u(x,\rho(x,y))$
is the inverse Fourier-Legendre transform of the  spectral functions
$\sigma_\ell(x)=\sqrt{f_\ell(x)}$, see Eq.~(\ref{u_spe}), so it suffices to estimate $\sigma_\ell(x)$.
We seek an estimator that is applicable to high-dimensional problems with $10^9$ degrees of freedom or more.

The outline of this section is as follows.
We start by motivating, in section \ref{sec_estm_bandpass_flt}, 
the application of a {\em multi-scale bandpass filter} to ensemble perturbations
with the intention to extract aggregated local spectra.
Then we explain how the (computationally intractable) likelihood of the raw data $\{\xi^{(k)}(x_i)\}$
(with $i=1,2,\dots, n_{\rm x}$ and $k=1,2,\dots, K$) can be 
approximated by a much simpler {\em composite likelihood} \citep{Lindsay,Reid}, see section \ref{sec_estm_lik}.
After that, we propose a two-stage estimator, section \ref{sec_estm_two_stage}. 
At the first stage, aggregated local spectra
are computed as sample variances of the bandpass filtered fields. 
At the second stage, the local spectra are disaggregated from these sample variances
for each grid point independently.
We explore two approaches to the disaggregation. 
The first one, based on a linear estimating equation, allows us to prove {\em consistency}
of the resulting model estimator, section \ref{sec_estm_consi}.
The second one relies on a {\em neural network} to estimate the local spectra, 
section \ref{sec_nrBayes}. Finally, the spatially varying convolution kernel is 
computed from the estimated local spectra (by substituting them into Eq.~(\ref{u_spe})) and used in the analysis.
The estimator's workflow is summarized in section \ref{sec_estm_workfl}.

Our approach can be regarded as a modification of the technique  by \citet{Priestley},
who, for each frequency $\omega_0$, applied 
a narrow bandpass filter centered at $\omega_0$ to a non-stationary time series and 
used the filter's mean-squared output
as the estimate of his evolutionary spectrum at the same frequency.
Another relevant reference is \citet[][chapter 11]{Marinucci}, who applied a wavelet filter to 
extract the variance spectrum of an isotropic random field on the sphere.

\subsection{Multiple spatial bandpass filters}
\label{sec_estm_bandpass_flt}

In the {\em stationary} case, it is straightforward to derive from Eq.~(\ref{xi_statio_spe}) that 
 the likelihood of the data, \ie the ensemble of random-field realizations $\{\xi^{(k)} (x)\}_{k=1}^{K}$,
 given the spectrum $\{f_\ell\}_{\ell=0}^{\ell_{\rm max}}$, depends on the data 
through the set of spectral-space single-wavenumber sample  variances
$\widehat v_\ell = \frac{1}{K}\sum_{k=1}^{K} \sum_{m=-\ell}^\ell |\widetilde\xi_{\ell m}^{(k)}|^2$. 
(Note that with $K=1$, the set of $\widehat v_\ell$ is the spherical analog of what is 
known in the time series literature as the {\em periodogram}.)
Therefore, the vector of the  single-wavenumber sample  variances 
$\widehat v_\ell$ (with $\ell=0,1,\dots,\ell_{\rm max}$)
is a sufficient statistic for the spectrum $f_\ell$.
This implies that  no information on the true spectrum is lost if we switch from the raw ensemble
to the set of  $\widehat v_\ell$.

In the locally stationary case, however, relying on the single-wavenumber sample  variances 
is not a good idea because isotropic spatial filters  that isolate individual spectral 
components have  non-local impulse response functions. 
Indeed, the filter's response function is the backward Fourier-Legendre transform of 
its spectral transfer function $H(\ell)$. If $H(\ell)$ 
is equal to one at the single wavenumber $\ell_0$ and zero otherwise, then
$h(\rho(x,y)) = 
\frac{2\ell_0+1}{4\pi} P_{\ell_0}(cos(\rho(x,y)))$ 
(see Eq.~(\ref{u_spe})). 
Due to the non-locality of $P_{\ell_0}(cos(\rho))$, the filtered signal, which is the 
convolution of the input signal with $h(\rho(x,y))$,  will contain significant
contributions from the whole global domain. This is not acceptable in the non-stationary case,
in which we seek primarily local contributions.

Therefore, instead of $\ell_{\rm max} +1$ filters that isolate individual spectral components, we introduce
a smaller number, $J$,  of {\em bandpass filters} ${\cal H}_j$ ($j=1,2,\dots,J$) 
that have broader spectral transfer functions $H_j(\ell)$. Broader transfer functions imply spatially 
localized impulse response functions, $h_j(\rho)$, as required with a locally stationary process.
Different $H_j(\ell)$ represent partly overlapping spectral {\em bands} that,
jointly, cover the whole spectral range.
Importantly,  the filtered signals $\varphi_{j}(x) = {\cal H}_j \xi(x)$
evaluated at a single grid point $x$ contain spatial information 
about the signal in question  $\xi(x)$ in a vicinity of $x$.
So, we can infer local spatial statistics of $\xi(\cdot)$ from the set of $\varphi_{j}(x)$ evaluated at a single point $x$.

On the other hand, the likelihood of $\xi(x)$ is computationally intractable 
because it depends on a huge number of parameters,
that is, the spectral functions  $\sigma_{\ell}(x)$ evaluated at all spatial grid points $x$.
In the estimator, we propose to replace the raw data $\xi(x)$
with the set of filtered fields $\varphi_{j}(x)$ and, crucially, switch from the likelihood of 
the whole spatial fields to the likelihood of 
$\boldsymbol\varphi(x) = \{\varphi_{j}\}_{j=1}^J(x)$ 
evaluated at each grid point $x$ individually.
As we show below in section \ref{sec_estm_lik}, the likelihood of $\boldsymbol\varphi(x)$ depends,
approximately, on  $\sigma_{\ell}(x)$ evaluated at the same  point. This reduces the global
estimation problem  to a series of local problems of much smaller dimensionality.

We note that broadening the filters' transfer functions 
reduces the spectral resolution of the estimator.
But, according to our fourth constraint (section \ref{sec_constr}),
the spatial spectra are smooth, so high spectral resolution is not needed.
For more details on the bandpass filters, see sections \ref{sec_estm_lik}, 
\ref{sec_expm_anls},  and \ref{sec_expm_LSEF}.

\subsection{Approximate likelihood of the filtered signals}
\label{sec_estm_lik}

Consider the application of the filter ${\cal H}_j$ to $\xi(x)$.
As shown in  \ref{App_accu_est}, if 
%
$L_{{\cal H}_j} \ll \Lambda$, 
%
(where $L_{{\cal H}_j}$ is the length scale of the $j$th bandpass filter's impulse response function
and $\Lambda$ the process's non-stationarity length scale introduced in section \ref{sec_LSM_locsta}),
then the spectral-space filtering can be approximately performed as if 
the random field $\xi(x)$ were stationary:
\begin {equation}
\label{phi1}
\varphi_{j}(x) = \sum^{\ell_{\rm max}}_{\ell=0} \sum^\ell_{m=-\ell}
   H_j(\ell) \,\sigma_{\ell}(x) \, \widetilde\alpha_{\ell m}  \, Y_{\ell m} (x) + \delta\varphi_{j}(x),
\end {equation}
where $\delta\varphi_{j}(x)$ is the discrepancy due to the non-stationarity, which has mean zero and variance 
$\Ex|\delta \varphi_j|^2  = O\left( L_{{\cal H}_j}^2 / \Lambda^2 \right)$ as 
$L_{{\cal H}_j}/\Lambda \to 0$, see Eq.~(\ref{LSM_S1_H_kappa2}).
The key point here is that, up to the small discrepancy,
the distribution of $\varphi_{j}(x)$ is determined by the spectral functions
$\sigma_{\ell}(x)$ evaluated at the same spatial point $x$ only. 

Specifically, given  $\boldsymbol\sigma(x) = \{\sigma_{\ell}(x)\}_{\ell=0}^{\ell=\ell{\rm max}}$,
the random vector 
$\boldsymbol\varphi(x) = \{\varphi_{j}(x)\}_{j=1}^J$  is  multivariate Gaussian with mean zero and covariances
$\Ex \varphi_{j}(x)\, \varphi_{j'}(x) = \sum^{\ell_{\rm max}}_{\ell=0} \frac{2\ell +1}{4\pi}
  H_j(\ell) H_{j'}(\ell) \,\sigma_{\ell}^2(x) + O(L_{{\cal H}_j} / \Lambda) +  O(L_{{\cal H}_{j'}} / \Lambda)$.
If the small $O(\cdot)$ contributions are neglected, this  fully determines its likelihood 
$p(\boldsymbol\varphi(x) \given \boldsymbol\sigma(x))$ for each grid point $x$.

For computational reasons, instead of relying on the exact likelihood of $\boldsymbol\varphi(x)$,
we neglect cross-correlations between different $\varphi_j(x)$ at the same grid point $x$.
Besides, considering  the likelihood grid point by grid point, we, 
effectively, neglect cross-correlations between different grid points as well.
As a result, we end up with the approximate Gaussian {\em composite likelihood} defined 
for each spatial grid point separately.
For the whole sample of $K$ ensemble members the  composite log-likelihood reads (up to a constant)
\begin {equation}
\label{CL}
c\ell[\boldsymbol\sigma(x)] 
  = -\frac{K}{2} \sum_{j=1}^J \left( \ln v_j(x) + \frac{d_j(x)}{v_j(x)} \right),
\end {equation}
where
$v_{j}(x) =\Var (\varphi_j(x) \given  \boldsymbol\sigma(x))$ are the {\em band variances} and
$d_j(x)$ are their sample counterparts, the {\em sample band variances}.
From Eq.~(\ref{phi1}), 
\begin {equation}
\label{vj}
v_{j}(x) = \utilde{v}_{j}(x) + \delta^{\rm nsta}_j(x),
\end {equation}
where
\begin {equation}
\label{vj2}
\utilde{v}_{j}(x) = \sum_{\ell=0}^{\ell_{\rm max}} \frac{2\ell +1}{4\pi}  H_j^2(\ell) \, \sigma_\ell^2(x)
\end {equation}
is the approximate band variance and $\delta^{\rm nsta}_j(x) = O({L_{{\cal H}_j}} / {\Lambda})$ 
is the approximation error.

According to Eqs.~(\ref{vj}) and (\ref{vj2}), the band variances $v_{j}(x)$ are, 
approximately, the aggregated local spectra $f_\ell(x) = \sigma_\ell^2(x)$. 
Jointly, $\{ v_{j}(x) \}_{j=1}^J$ bear all the information on the (smooth) local spectrum
we wish to infer from the sample (up to the small discrepancy $\delta^{\rm nsta}_j$).
On the other hand, the sample band variances $d_j(x)$ summarize information on the local spectrum
contained in the raw sample $\{\xi^{(k)}(x) \}_{k=1}^K$  
(up to the sampling error: $d_j(x) = v_j + \delta^{\rm samp}_j$).
Note that $v_{j}(x)$ and  $d_j(x)$ are the locally 
stationary counterparts of the stationary spectrum $f_{\ell}$ and  periodogram $\widehat v_\ell$,
respectively, see section \ref{sec_estm_bandpass_flt}.

\subsection{A two-stage estimator}
\label{sec_estm_two_stage}

Equations (\ref{CL})--(\ref{vj2}) suggest the following two-stage estimator applied point-wise,
for each spatial grid point $x$ independently:
\begin{enumerate}
\item
\label{list_stage1}
Compute the sample band variances $d_j(x)$:
	\begin{enumerate}
	\item
Bandpass filter $K$ ensemble perturbations,
$\xi^{(k)} (\cdot)$, $k=1,2\dots,K$. The result is the $JK$ filtered fields 
$\varphi_j^{(k)}(\cdot)$, $j=1,2\dots,J$.
	\item
At each spatial grid point $x$, compute sample variances of $\varphi_j^{(k)}(x)$, that is,
the sample band variances  $d_j(x)$.
	\end{enumerate}

\item
\label{list_stage2}
Recover $\boldsymbol\sigma(x)$ from the set of sample band variances ${\bf d}(x) = \{ d_j(x) \}_{j=1}^J$. 
We consider two ways of doing so.
	\begin{itemize}
	\item
The first approach relies on linearity of Eq.~(\ref{vj2}) with respect to the local spectrum
$f_\ell(x) =  \sigma_\ell^2(x)$ and neglect of $\delta^{\rm nsta}_j(x)$. 
We rewrite Eqs.~(\ref{vj}) and (\ref{vj2}) as 
\begin {equation}
\label{vf}
{\bf v}(x) \approx \boldsymbol\Omega \,{\bf f}(x),
\end {equation}
where 
${\bf v}(x) = \{ v_j(x) \}_{j=1}^J$,
${\bf f}(x)=\{f_\ell(x)\}^{\ell_{\rm max}}_{\ell=0}$, and
$\boldsymbol\Omega$ is the $J$ by $\ell_{\rm max}+1$ matrix with the known entries 
$(\boldsymbol\Omega)_{j\ell} = \frac{2\ell +1}{4\pi}  H_j^2(\ell)$.

Equations~(\ref{CL}) and (\ref{vf}) imply that the composite likelihood
$c\ell[\boldsymbol\sigma(x)]$ is approximately maximized at any vector ${\bf f}(x)$ that satisfies
\begin {equation}
\label{estmEq}
{\bf d}(x) = \boldsymbol\Omega \,{\bf f}(x).
\end {equation}
This estimating equation is underdetermined and thus needs to be regularized.
In section \ref{sec_estm_consi}, we do so by imposing a deterministic smoothness constraint and show that
the error of the resulting solution vanishes in the large sample limit
so that the whole estimator is consistent.

	\item
The second approach is to 
rely on the available unapproximated nonlinear data generation (or `forward') model
$\boldsymbol\omega$ such that 
\begin {equation}
\label{nlinInvProbl}
{\bf d}(x) = \boldsymbol\omega(\boldsymbol\sigma(\cdot), x).
\end {equation}
In section \ref{sec_nrBayes}, we build an optimized neural-network based mapping 
${\bf d}(x) \mapsto \boldsymbol\sigma(x)$ that approximates the classical Bayes estimator.

	\end{itemize}

In the next two subsections, we elaborate on the two implementations of stage 2, 
one based on Eq.~(\ref{estmEq}) and the other on Eq.~(\ref{nlinInvProbl}).

\end{enumerate}

\subsection{Consistency of the linear estimator}
\label{sec_estm_consi}

To solve the underdetermined Eq.~(\ref{estmEq})
we employ the deterministic regularization technique 
known as `spectral cut-off', \citep[e.g.][]{Cavalier}.
Specifically, we impose a smoothness constraint
that requires  the local spectrum to be band-limited as a function of log-wavenumber, see below.

To formulate the smoothness constraint, we switch from wavenumbers $\ell$ to log-wavenumbers
$\vartheta(\ell) = \pi \log (\ell+1) / \log (\ell_{\rm max}+1)$.
We set $\vartheta(-\ell) = \vartheta(\ell)$ so that $\vartheta$ ranges from $-\pi$ to $\pi$.
Then, we let $F(\vartheta(\ell)) = f_\ell$. We also assume
that the spatial resolution is large enough and the spectra are smooth enough 
to treat $F(\vartheta)$ as a function of continuous argument
on the unit circle.
\begin{definition}
\label{def_smoo}
We say that the spectrum $f_\ell$ is {\em smooth with the bandwidth $M$} if 
 $F(\vartheta)$ is a {\em band-limited function}:
\begin {equation}
\label{FFk}
F(\vartheta) = \sum_{m=-M}^M \widetilde F_m \e^{\i m \vartheta},
\end {equation}
where $\widetilde F_m$ is the the Fourier transform of $F(\vartheta)$ and 
$M$ is a positive integer.
\end{definition}
%
The smoothness constraint in this section is the assumption that the spectrum is
{\em smooth with the bandwidth $M$} (the lower $M$, the smoother $F(\cdot)$). 
Since $F(\vartheta)$ is an even real-valued function, it follows from
Eq.~(\ref{FFk})  that the spectrum becomes parameterized with the vector, ${\bf p}$, of
$M+1$ real-valued parameters:
${\bf f}= \boldsymbol\Psi{\bf p}$ with the known  $(\ell_{\rm max}+1)\times (M+1)$ matrix $\boldsymbol\Psi$.
Therefore, we can select $J=M+1$ bandpass filters so that the number of unknowns becomes equal to the number
of data (the sample band variances ${\bf d}$). Then, Eq.~(\ref{estmEq}) reduces
to  a system of linear algebraic equations: 
\begin {equation}
\label{linInvProbl2}
{\bf d} = \boldsymbol\Omega\boldsymbol\Psi {\bf p}.
\end {equation}
The solution to this equation is subject to both sampling error
$\boldsymbol\delta^{\rm samp}(x) = \{ \delta^{\rm samp}_j(x) \}_{j=1}^J$  
and non-stationarity error 
$\boldsymbol\delta^{\rm nsta}(x) = \{ \delta^{\rm nsta}_j(x) \}_{j=1}^J$.
In  \ref{App_consi} we prove that their impact on the estimated local spectrum
results in the estimation error that can be made arbitrarily small by increasing the sample size
and appropriately increasing spectral widths of the filters. More specifically, we prove
the following 
\begin{proposition}
\label{prop_consi}
Let the local spectrum be smooth in terms of Definition \ref{def_smoo} with the bandwidth $M$.
Then, for every sample size $K$, there  exist $M +1$  bandpass filters such that the solution to 
Eq.~(\ref{linInvProbl2}) yields vanishing mean absolute error in the local spectrum as $K \to\infty$.
The  spectral widths of the filters (in terms of log-wavenumbers) 
grow with the increasing sample size $K$ as $K^\gamma$ for some $\gamma > 0$.
\end{proposition}
Due to Markov's inequality, Proposition \ref{prop_consi} entails that the estimator is consistent.

Two points are worth noting. 
First, while proving consistency we found that the widths of the 
bandpass filters' transfer functions are to be selected  
depending on the ensemble size:  the bigger the ensemble,
the broader bandpass filters' spectral  transfer functions can and should be taken.
Broader  transfer functions imply more localized filters'
impulse response functions and therefore less non-stationarity error $\delta^{\rm nsta}_j$ in $d_{j}(x)$.
This allows the estimator to extract increasingly detailed
spatially non-stationary structures, resulting in more accurately estimated local spectra.
Second, this conclusion has important consequences not just for the theoretical proof of 
consistency but also for a more practical implementation of stage 2 of the estimator 
(presented in the next subsection), where such a conclusion could not be derived analytically.

\subsection{Neural Bayes estimator of local spectrum from sample band variances}
\label{sec_nrBayes}

We recall that the second implementation of stage 2 of the estimator,
see section \ref{sec_estm_two_stage}, abandons the linear approximation
made in the previous subsection and relies on  Eq.~(\ref{nlinInvProbl}. 
This makes the `forward model' more accurate and reduces the noise level --- but at the expense of
rendering the problem nonlinear and high-dimensional.
Indeed, the exact expression for the filtered process $\varphi_j(x)$ is
\begin {equation}
\label{phi_exact}
\varphi_j(x) = \int_D h_j(\rho(x,y))  \xi(y)  \,\d y = 
  \sum^{\ell_{\rm max}}_{\ell=0} \sum^\ell_{m=-\ell} \widetilde\alpha_{\ell m}
     \int_D h_j(\rho(x,y)) \,\sigma_\ell(y) \, Y_{\ell m}(y)   \,\d y,
\end {equation}
where Eq.~(\ref{LSM}) was used. From Eq.~(\ref{phi_exact}), it follows that
the respective  band variance $v_j(x) = \Ex (|\varphi_j(x)|^2 \given \boldsymbol\sigma(.))$
depends not only on $\sigma_\ell^2(x)$ (\ie evaluated at the same point $x$ at which $\varphi_j$
is evaluated),  but also on the products 
$\sigma_\ell(y) \, \sigma_\ell(y')$ for all pairs of spatial points $(y,y')$.
This justifies writing the data model as
${\bf d}(x) = \boldsymbol\omega(\boldsymbol\sigma(\cdot), x)$, as in
Eq.~(\ref{nlinInvProbl}).
In high dimensions, this also implies that the likelihood of ${\bf d}(x)$ is numerically intractable.
Therefore we build an  estimator of $\boldsymbol\sigma(x)$
that does not require evaluations of the likelihood function.

We start with postulating that the estimator  is of the form 
\begin {equation}
\label{NNe}
\widehat{\boldsymbol\sigma}(x) = \boldsymbol\Phi({\bf d}(x), \boldsymbol\psi),
\end {equation}
where $\boldsymbol\Phi = (\Phi_0, \Phi_1, \dots, \Phi_{\ell_{\rm max}})$ 
is a known vector function of its arguments and 
$\boldsymbol\psi$ is the unknown vector of the estimator's parameters. 

We place a prior distribution, $p(\boldsymbol\sigma(\cdot))$, over 
the unknown spectral functions $\sigma_\ell(\cdot)$. 
Note that $p(\boldsymbol\sigma(\cdot))$ is a hyperprior in the context of ensemble filtering.
The hyperprior $p(\boldsymbol\sigma(\cdot))$ encodes the four constraints of our spatial model, 
see section \ref{sec_constr}.

To optimize the estimator, that is, to find the best $\boldsymbol\psi$,  we introduce a {\em loss function}
chosen to be quadratic: 
\begin {equation}
\label{loss}
{\cal L}[\widehat{\boldsymbol\sigma}(\cdot), \,\boldsymbol\sigma(\cdot)] = \frac{1}{|D|} \int_D
  \sum_{\ell=0}^{\ell_{\rm max}}  \frac{2\ell +1}{4\pi} 
 \left(\widehat{\sigma_\ell}(x) - \sigma_\ell(x) \right)^2 \,\d x
\end {equation}
(where $|D|$ is the area of the analysis domain $D$).
Aiming at the Bayes point estimator  \citep[e.g.][Chapter 4]{Lehmann},
we minimize the Bayes risk, that is, the expectation of the loss function with respect 
to the joint distribution of the parameter
$\boldsymbol\sigma(\cdot)$ and the data ${\bf d}(\cdot)$:
\begin {equation}
\label{psi}
\widehat{\boldsymbol\psi} = \argmin_{\boldsymbol\psi} \Ex \,
{\cal L}[\boldsymbol\Phi({\bf d}(\cdot), \boldsymbol\psi), \;\boldsymbol\sigma(\cdot)],
\end {equation}
getting the mean-square optimal estimator's parameters $\boldsymbol\psi$ and thus the optimal
estimator defined by Eq.~(\ref{NNe}).

As the likelihood function is  numerically intractable, we
rely on a Monte-Carlo approximation to solve the optimization problem Eq.~(\ref{psi}). 
In doing so, we assume that, first, there is 
a sample $\{\boldsymbol\sigma^i(\cdot)\}_{i=1}^I$ ($I$ is its size) 
drawn from the hyperprior distribution  $p(\boldsymbol\sigma(\cdot))$. 
Second, we exploit the fact that the data model
$ \boldsymbol\omega(\boldsymbol\sigma(\cdot),x)$ is  {\em generative}: given $\boldsymbol\sigma^i(\cdot)$,
it can simulate the data ${\bf d}^{i}(x_r)$, that is, 
$J$ sample band variances at $R$ randomly and uniformly selected grid points $x_r$, where $r=1,2,\dots,R$.
Since, by construction, 
$\boldsymbol\sigma^i(\cdot) \sim p(\boldsymbol\sigma(\cdot))$ and
${\bf d}^{i}(\cdot) \sim p({\bf d}(\cdot) \given \boldsymbol\sigma^i(\cdot))$,
we can approximate the expectation in Eq.~(\ref{psi})) and the integral in Eq.~(\ref{loss}) 
using Monte-Carlo:
\begin {equation}
\label{NlinInvProbl}
\frac{1}{n_{\rm train}} \sum_{i=1}^I \sum_{r=1}^R  
\sum_{\ell=0}^{\ell_{\rm max}}  \frac{2\ell +1}{4\pi} 
 \left(\Phi_\ell({\bf d}^{i}(x_r), \boldsymbol\psi) - \sigma^i_\ell(x_r) \right)^2,
\end {equation}
where
$n_{\rm train} = IR$ is the size of the training sample.

In this study, we choose a neural network as the parametric estimator 
$\boldsymbol\Phi({\bf d}(x), \boldsymbol\psi)$ with
$\boldsymbol\psi$ being the collection of the net's weights and biases.
If 
(i) the training sample  $\{\boldsymbol\sigma^i(\cdot)\}_{i=1}^I$ is large and representative
of the hyperprior $p(\boldsymbol\sigma)$,
(ii) the neural network is expressive enough, and 
(iii) the  network optimizer is capable of 
finding a good proxy to the global minimum in Eq.~(\ref{NlinInvProbl}), 
then the resulting estimator is a good approximation to the Bayes estimator.
In this way, we obtain a {\em neural Bayes estimator} \citep{Zammit2024}.
Technical details on the  neural net are given in \ref{App_NN}.

The above neural-Bayes based 
disaggregation of the local spectrum from the set of sample band variances
makes the technique applicable to high-dimensional problems. This is because 
(i) the estimator is local in space, that is, for each spatial point $x$, 
it maps ${\bf d}(x)$ to $\widehat{\boldsymbol\sigma}(x)$,
(ii) the resulting inference is {\em amortized}: no retraining is needed when new data arrive
(unless there are changes in the hyperprior  $p(\boldsymbol\sigma(\cdot))$ or in the ensemble size, 
or in the bandpass filters),
(iii) the computations can be performed at different grid points in parallel. 

It is worth reiterating that the estimator 
$\widehat{\boldsymbol\sigma}(x) = \boldsymbol\Phi({\bf d}(x), \boldsymbol\psi)$
is trained only once, that is, its parameters $\boldsymbol\psi$ are optimized during the training
and then kept fixed. The application of the estimator 
to fresh data  ${\bf d}(x)$ requires no additional optimization.
The neural network just transforms its input ${\bf d}(x)$ to the estimate 
$\widehat{\boldsymbol\sigma}(x)$.

At the final stage of the estimator, we substitute 
$\widehat\sigma_\ell(x)$ into Eq.~(\ref{u_spe}), getting the 
estimated spatially varying convolution kernel $\widehat u(x, \rho)$, which, in turn,  is
used to compute the $\widehat{\bf W}$ matrix, the key ingredient of the
analysis algorithm, see section \ref{sec_LSEF_anls}.
The resulting spatial covariances, being determined by both the hyperprior (`climatological')
distribution of the local spectra and the current ensemble, are non-stationary in time 
as well as in space.
The uncertainty of the estimate is not taken into account in this study.

\subsection{Summary of the estimator's workflow}
\label{sec_estm_workfl}

The practical estimation procedure consists of the following steps.
\begin{itemize}
\item
Ensemble perturbations are bandpass filtered. 
\item
Sample variances of the 
filtered fields are computed at each grid point.
\item
Local spectra are inferred from the sample band variances using a neural network
grid point by grid point.
\item
The spatially varying convolution kernel is computed from the estimated local spectra.
\end{itemize}

\section {LSEF: A Locally Stationary Ensemble Filter}
\label{sec_LSEF_anls}

The approach taken in this study is to use a spatial random field model to regularize 
the prior covariance matrix in sequential ensemble filtering.
We call the resulting regularized EnKF  the
Locally Stationary Ensemble Filter (LSEF).
Its forecast step coincides with that of the stochastic EnKF.
The only difference between the LSEF's  and the traditional-EnKF's analyses 
is how the proxy $\widehat{\bf B}$ to the true prior covariance matrix is defined.
In this way, we can isolate the impact of our spatial-model-based regularization approach.

Testing and validation of our EnKF regularization technique can well be performed in the {\em linear} setting,
avoiding unnecessary complications related to nonlinearities of observation and forecast models.
For this reason, in the sequel we confine ourselves to the linear filtering setup.
\ref{App_lin_seq_flt} reviews optimal and sub-optimal, ensemble and non-ensemble,
 time-discrete linear filters that are used in the numerical experiments presented below.

In the LSEF analysis scheme, we postulate that the forecast error obeys the
Locally Stationary Convolution Model. The model, or more specifically the
spatial convolution kernel $u(x,\rho(x,y))$, is estimated  from the forecast ensemble and a hyperprior distribution
as detailed in section \ref{sec_LSM_estm}. 
With the estimated  convolution kernel $\widehat u(x,\rho(x,y))$ in hand, 
we build the space-discrete approximation of Eq.~(\ref{osc2}):
\begin {equation}
\label{osc2dm}
\boldsymbol\xi  = \widehat{\bf W} \boldsymbol\alpha,
\end {equation}
where $\boldsymbol\xi$ is the vector that represents the random field $\xi(x)$
evaluated on the spatial grid,
$\boldsymbol\alpha \sim \mathcal{N}({\bf 0,I})$ and $\widehat{\bf W}$ is a matrix with the entries
\begin {equation}
\label{wij}
\widehat w_{ii'} = \widehat u(x_i, \rho(x_i, x_{i'})) \,\sqrt{\Delta x_{i'}},
\end {equation}
where $\Delta x_{i'}$ is the area of ${i'}$th grid cell.
Note that Eqs.~(\ref{osc2dm}) and (\ref{wij}) approximate the integral in Eq.~(\ref{osc2}) using the simplest
rectangle rule, more accurate approximations will be considered elsewhere.
Equation (\ref{osc2dm}) entails that
\begin {equation}
\label{BW_}
\widehat{\bf B} = \widehat{\bf W} \widehat{\bf W}^\TT.
\end {equation}
We note that the representation of the forecast-error covariance matrix $\widehat{\bf B}$ in the `square-root' form, Eq.~(\ref{BW_}),
is common in data assimilation practice because it provides efficient preconditioning of the analysis equations
\citep[e.g.][]{Asch}.
The need for preconditioning follows from Eq.~(\ref{K}) because
the matrix ${\bf B}^{-1} + {\bf H}^\TT {\bf R}^{-1}{\bf H}$ to be inverted
is, normally, ill-conditioned.
Substituting Eq.~(\ref{BW_}) into Eq.~(\ref{K}) and rearranging the terms yields
\begin {equation}
\label{gainW} 
\widehat{\bf K} = \widehat{\bf W} \left({\bf I} + 
   \widehat{\bf W}^\TT{\bf H}^\TT {\bf R}^{-1}{\bf H}\widehat{\bf W} \right)^{-1} 
    \widehat{\bf W}^\TT{\bf H}^\TT {\bf R}^{-1}.
\end {equation}
Now the matrix to be inverted, 
${\bf I} + \widehat{\bf W}^\TT{\bf H}^\TT {\bf R}^{-1}{\bf H}\widehat{\bf W}$, 
is well conditioned (as a sum of the identity matrix and a non-negative definite matrix).
This is how the LSEF analysis is done, that is, without explicitly computing $\widehat{\bf B}$.
The LSEF analysis assimilates all available observations at once,
as this is done  in variational data assimilation schemes \citep[e.g.][]{Asch}.

The representation of the covariance matrix $\widehat{\bf B}$ as $\widehat{\bf W} \widehat{\bf W}^\TT$ also
guarantees  positive definiteness of $\widehat{\bf B}$ for any $\widehat{\bf W}$.
This property can be utilized to perform {\em thresholding} of the $\widehat{\bf W}$ matrix, \ie
nullifying its small in modulus entries to make it sparse and
facilitate fast computations without jeopardizing positive definiteness of the implied covariance matrix.

\section{Numerical experiments with static analyses}
\label{sec_expm_anls}

In this section, we experimentally examine the above LSEF {\em analysis} technique in the setting with 
known synthetic `truth' (ground truth) that obeys the Locally Stationary Convolution Model on the sphere.
Besides, the hyperprior distribution of the model parameters was known in these experiments as well, so 
the neural Bayes estimator could be trained using the perfect training sample.
This setting allowed us to reveal the best 
possible performance our approach can provide.

We compared deterministic analyses produced by the following schemes:
	\begin{itemize}
	\item
True-B: the optimal analysis (which has access to the true ${\bf B}$  matrix).
	\item
LSEF-B: the new analysis proposed in this paper (with the neural Bayes estimator).
	\item
Mean-B:  analysis with time-mean ${\bf B}$ (known in geosciences as 3D-Var). 
	\item
EnKF-B: stochastic EnKF with tuned localization.
	\item
Hybrid-B: analysis with the half-sum of the time-mean ${\bf B}$ and EnkF's  ${\bf B}$
(known as EnVar). 
	\end{itemize}

See section \ref{sec_LSEF_anls} for the description of the LSEF analysis and 
 \ref{App_anls_nonens} and \ref{App_anls_ens}
for the description of the other (reference) analyses.
The quality of the analysis ensemble was not tested, it will be  assessed 
in sequential filtering experiments presented in section \ref{sec_expm_LSEF}.

\subsection {Experimental methodology}
\label{sec_expm_meth}

\begin{figure}[ht]
\begin{center}
   { 
   \scalebox{0.63}{ \includegraphics{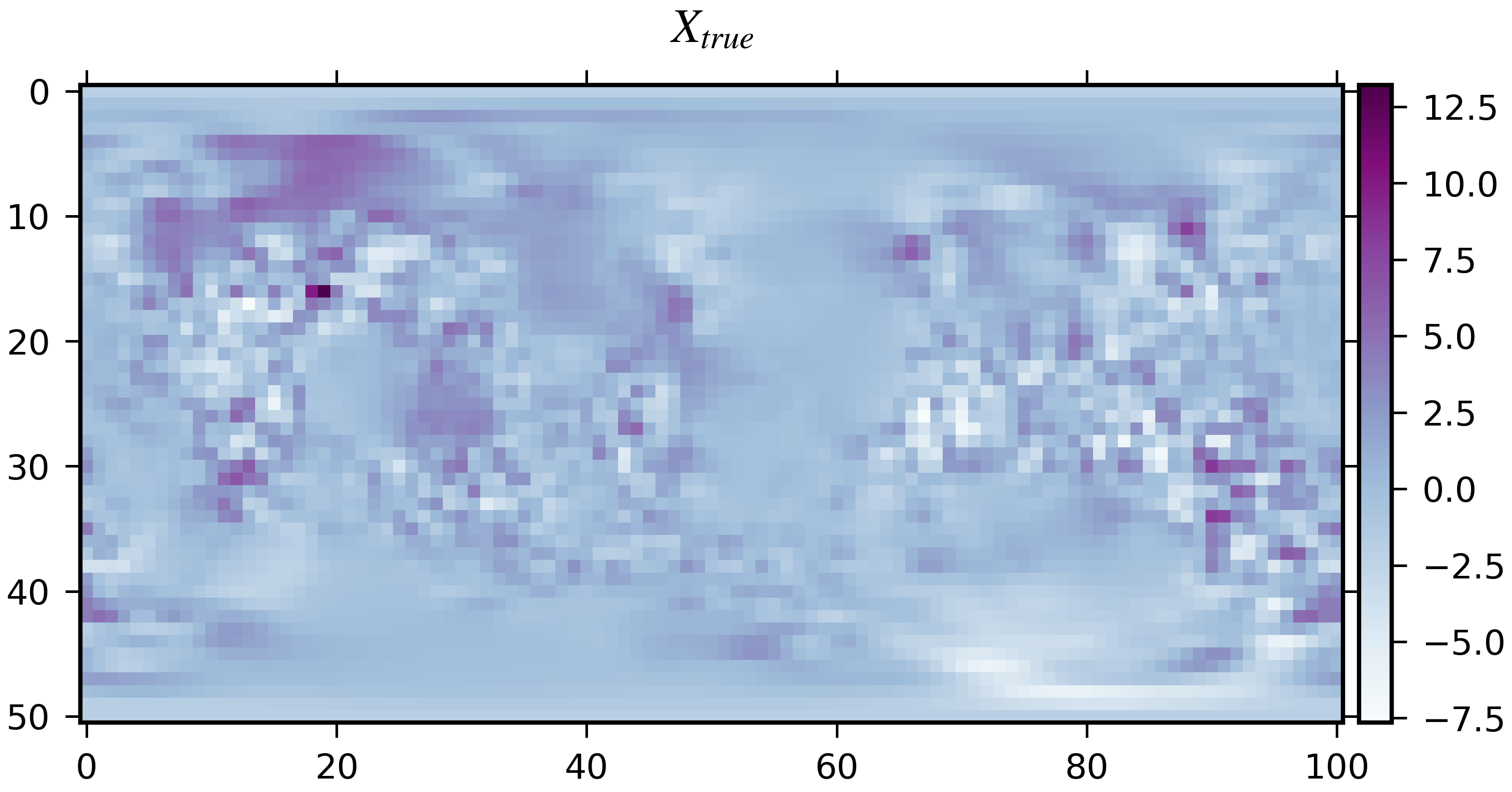}}
    \scalebox{0.63}{ \includegraphics{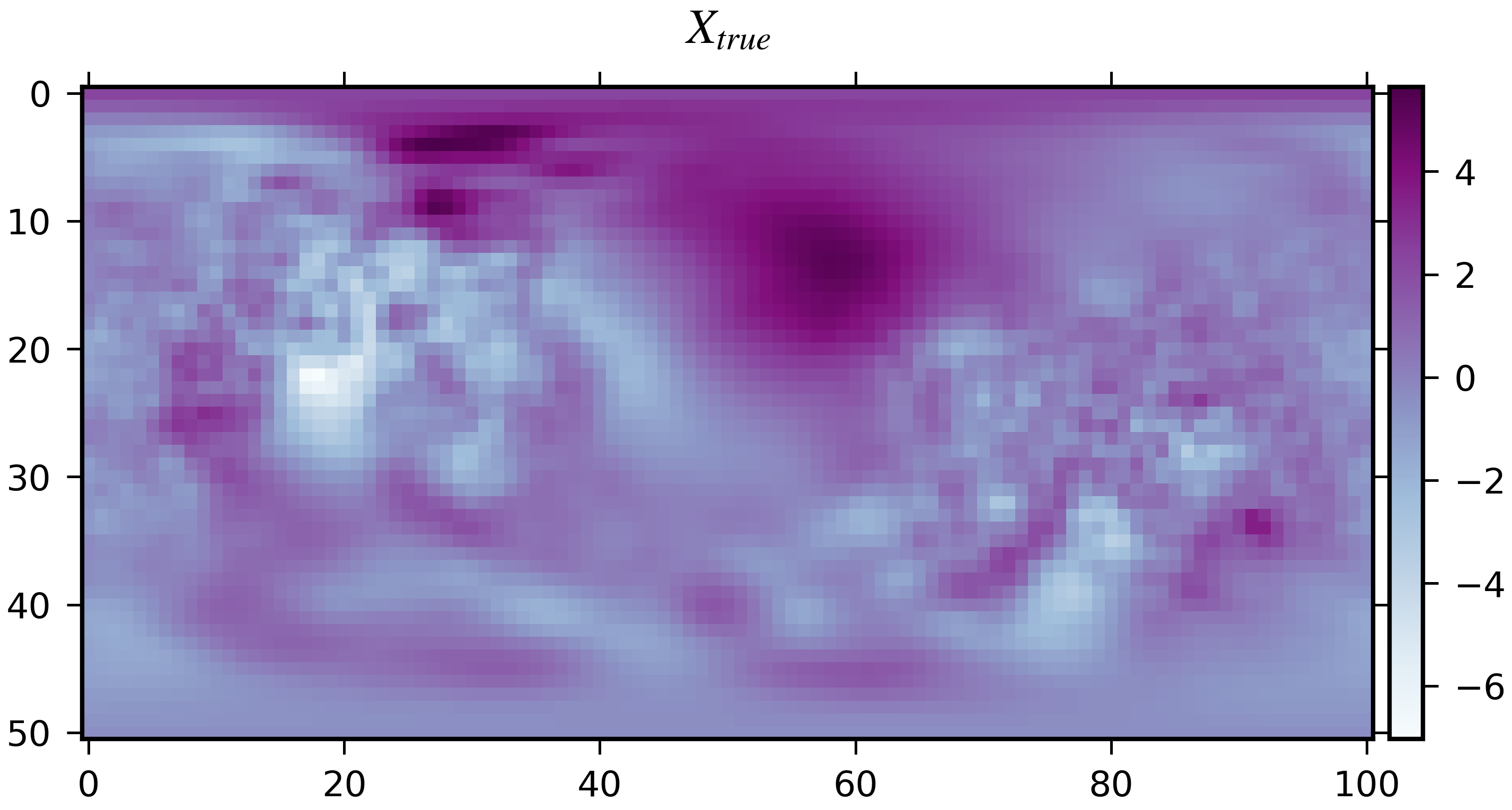}}
       }
\end{center}
  \caption{Two realizations of the random field produced by the parametric model of `truth'. 
   Left: $\mu_{\rm NSL}=1$. Right: $\mu_{\rm NSL}=3$. 
         x-axis: longitude index,        
         y-axis: co-latitude index.}
\label{Fig_xi}
\end{figure}

Pseudo-random realizations of the  `true' spectral functions $\sigma_\ell(x)$ 
(which, we recall, are non-negative square roots of the local spectra $f_\ell(x)$) and the 
respective `true' background error field $\xi(x)$  were generated following the
Parametric Locally Stationary Convolution Model of  `truth' described in
\ref{App_pLSM}.
To give the reader an impression of the ground truth fields, we
show in Fig.~\ref{Fig_xi} two pseudo-random realizations\footnote{
We used SHTools \citep{Shtools} to generate pseudo-random fields on the sphere.} of $\xi$
for two different values of the non-stationarity length parameter  
$\mu_{\rm NSL}$, which is defined in \ref{App_true_LSM_pre} as the ratio of the 
non-stationarity length scale $\Lambda_{\rm NSL}$ and the median process length scale $L_{\rm med}$.
The left panel of Fig.~\ref{Fig_xi} shows the field with the small $\mu_{\rm NSL}=1$, that is,
with the strong spatial non-stationarity.
The left panel of Fig.~\ref{Fig_xi} displays a realization of the random field with the default (moderate)
 non-stationarity, $\mu_{\rm NSL}=3$.
In both panels of Fig.~\ref{Fig_xi}, one can see a 
substantial degree of spatial non-stationarity, both in the field's magnitude
and  length scale.
With $\mu_{\rm NSL}=1$, Fig.~\ref{Fig_xi}(left),
areas where the local length scale of $\xi$ is small (seen as patchy spots) and 
where the local length scale is large (smooth spots) 
are both seen to be smaller in size compared with those in the right panel.
This reflects a smaller non-stationarity length scale with $\mu_{\rm NSL}=1$ than with 
 $\mu_{\rm NSL}=3$, as expected.

The true background-error vector $\boldsymbol\xi$ 
was computed by evaluating $\xi(\cdot) \given \sigma$ on the analysis grid
(defined below in section \ref{sec_expm_anls_setup}).
We assumed (without loss of generality) that  ${\bf x}^{\rm f}=0$ 
so that the simulated ground truth was ${\bf x} = {\bf x}^{\rm f} - \boldsymbol\xi = - \boldsymbol\xi$
and the ensemble members were 
${\bf x}^{{\rm fe}(k)} = {\bf x}^{\rm f} - \boldsymbol\xi^{(k)} = -\boldsymbol\xi^{(k)}$
(see \ref{App_lin_seq_flt} for notation).

Having computed the true $\sigma_\ell(x)$, we used Eq.~(\ref{u_spe}) 
to compute the true kernel $u(x, \rho)$.
After that, we built the true  ${\bf W}$ matrix following Eq.~(\ref{wij}) and applied it 
to independent realizations of the white noise $\boldsymbol\alpha$
to generate the  background-error vector $\boldsymbol\xi$
and the ensemble perturbations $\boldsymbol\xi^{(k)}$, see Eq.~(\ref{osc2dm}). 
The true covariance matrix (used by the optimal True-B analysis) was
${\bf B} = {\bf W} {\bf W}^\TT$.

With the truth ${\bf x} =  - \boldsymbol\xi$ in hand, we generated  
observations ${\bf x}^{\rm obs}$ following Eq.~(\ref{obs}),
assimilated them using the five analysis techniques listed above, and 
compared the resulting deterministic analyses with the truth.

\subsection {Details of the analysis schemes}
\label{sec_expm_anls_details}

The four sub-optimal analysis schemes differed 
only in the way the prior covariance matrix 
was specified. 
In the EnKF analysis, the localization matrix 
was specified using the popular Gaspari-Cohn correlation function  \citep{Gaspari}, whose length scale
was tuned to obtain the best performance for each configuration of the  model of truth and for each ensemble size
separately.

In the Mean-B analysis, the static prior covariance matrix 
${\bf B}_{\rm static}={\bf B}_{\rm mean}$  was computed
for each configuration of the  model of truth as follows.
We generated $n_{\rm repl}=33$ replicates of the true spectral functions $\sigma_\ell(x)$.
Given each replicate, $\sigma_\ell^{(i)}(x)$, we generated a sample of
$n_{\rm sample}=10$ realizations of $\xi\given \sigma$: $\xi^{(i,k)}(\cdot)$, where
$i=1,2,\dots, n_{\rm repl}$ and $k=1,2,\dots, n_{\rm sample}$.
Then, we expanded each realization in spherical harmonics, getting the
spectral coefficients $\widetilde\xi^{(i,k)}_{\ell m}$. For each $\ell$, we summed
$|\widetilde\xi^{(i,k)}_{\ell m}|^2$ over $m$ in the range from $-\ell$ to $\ell$, divided by $4\pi$, and
averaged over $i$ and $k$, obtaining a sample estimate of the {\em mean variance spectrum} of $\xi$.
From the mean variance spectrum, we computed the mean (stationary in space) covariance function and
used it to build ${\bf B}_{\rm mean}$.

In the Hybrid-B analysis,  a convex combination
of the static and localized sample covariance matrices was used:
$\widehat{\bf B}_{\rm hybr} = (1-w_{\rm e}) {\bf B}_{\rm mean} + w_{\rm e} \widehat{\bf B}_{\rm EnKF}$,
where $w_{\rm e}=0.5$ was found to be nearly optimal and used in all experiments.

\begin{figure}[h]
\begin{center}
   { 
   \scalebox{0.55}{ \includegraphics{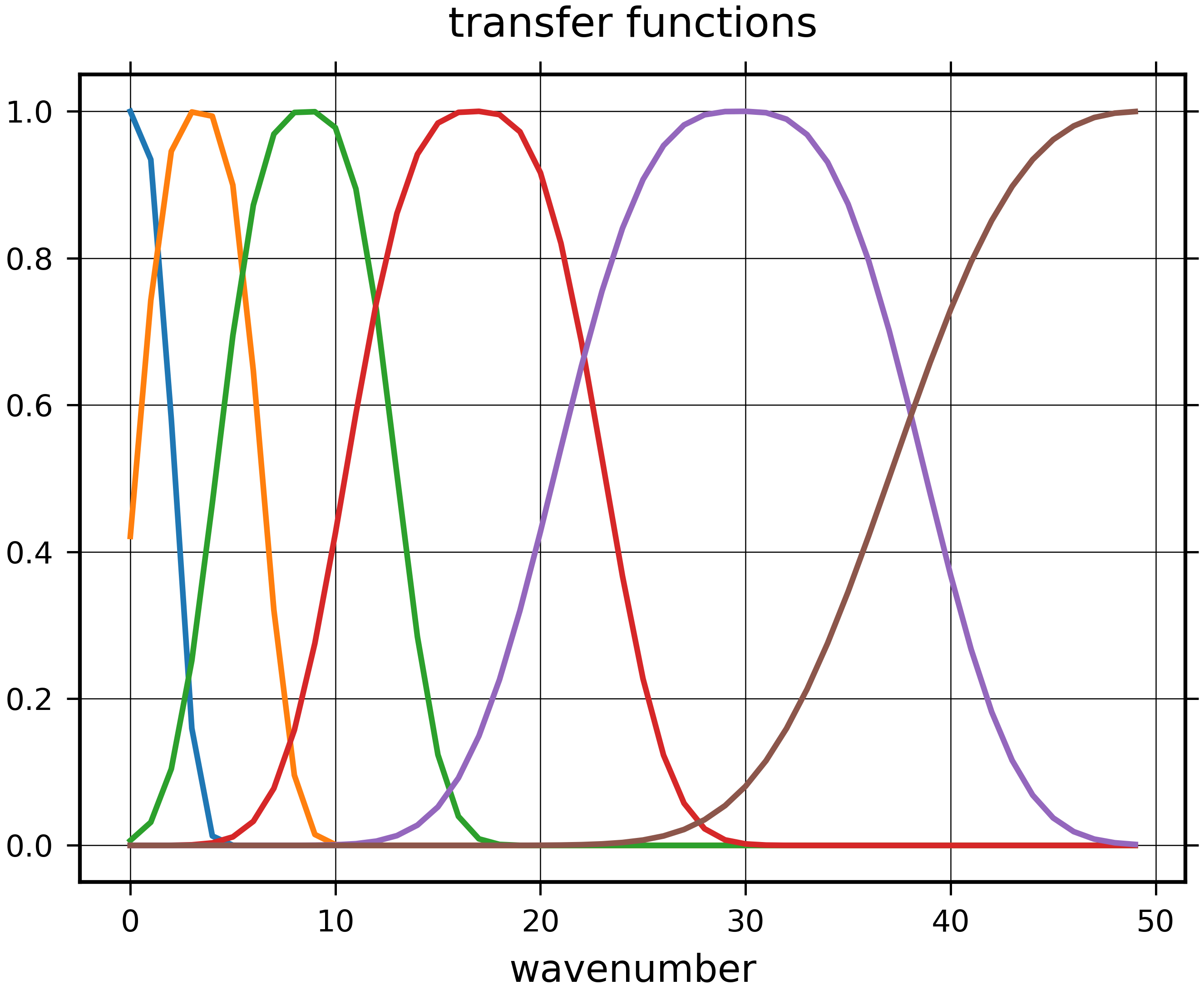}}
   \scalebox{0.55}{ \includegraphics{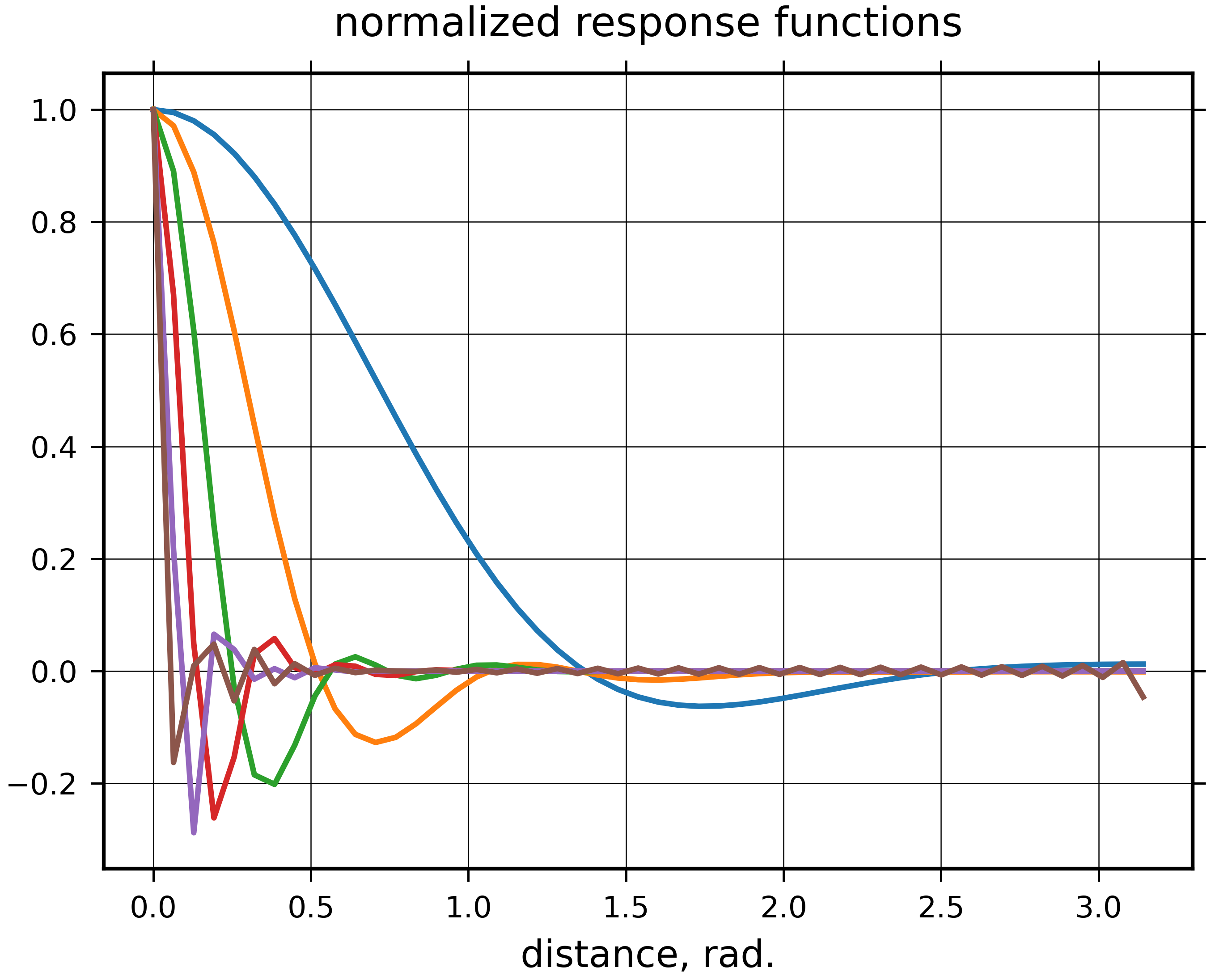}}
       }
\end{center}
  \caption{Spectral transfer functions $H_j(\ell)$ (left)  and their respective
        normalized impulse response functions $h_j(\rho)$ (right)
        for  six bandpass filters.}
\label{Fig_spaFlt}
\end{figure}

In LSEF-B, the bandpass filters' spectral transfer functions were specified as 
\begin {equation}
\label{Hj}
H_j(\ell) = \exp \left(-\left|\frac{\ell-\ell^{\rm c}_j}{\Delta_{{\cal H}_j}} \right|^{q_{\rm shape}} \right),
\end {equation}
where $\ell^{\rm c}_j$ is the central wavenumber of the $j$th filter,
$\Delta_{{\cal H}_j}$ is the the half-width parameter,
and $q_{\rm shape}$ is the shape parameter.
In the experiments described below, we took ${q_{\rm shape}}=3$ (${q_{\rm shape}}=2$ also worked well). Both
$\ell^{\rm c}_j$ and ${\Delta_{{\cal H}_j}}$ were taken exponentially growing with the band index  $j$.
Figure~\ref{Fig_spaFlt} shows the spectral transfer functions and the respective 
impulse response functions
for six bandpass filters used in the experiments presented  in this section.
The spatial filtering was performed in spectral space.

%

\subsection {Experimental setup}
\label{sec_expm_anls_setup}

The default setup of the parametric model of truth is 
specified in \ref{App_true_LSM_hyper}.

The maximal wavenumber was $\ell_{\rm max}=50$.
The analysis grid was of the regular latitude-longitude type  with $\ell_{\rm max}+1$ latitudes
 (so that the mesh size in the latitudinal direction was $\Delta x = \pi /\ell_{\rm max}$)
and $2\ell_{\rm max}$ longitudes. 
Note that according  to the spherical sampling theorem by \citet[][Theorem 3]{Driscoll},
a band-limited (with the bandwidth $\ell_{\rm max}$) function defined on the sphere can be losslessly
represented on the regular latitude-longitude grid that has $n_{\rm lat}=2\ell_{\rm max}$
latitudes and $n_{\rm lon}=2 \ell_{\rm max}$ longitudes.
With our decaying spectra, reducing $n_{\rm lat}$ to $\ell_{\rm max} +1$  led to negligible errors.

Observations were located at randomly and independently sampled grid points  
selected with probabilities proportional to the spherical surfaces of the respective grid cells. 
The number of observations was $n_{\rm obs}=\frac12 n_{\rm x}$
(recall that $n_{\rm x}$ stands for the total number of grid points,  that is, the length 
of the state vector ${\bf x}$). 
The observation-error variance was equal to the median background-error
variance (we tried other choices and obtained qualitatively very similar
results, not shown).

The ensemble size was between 5 and 80 with the default 20.

\subsection {Training sample}
\label{sec_expm_anls_train}

Training  the neural network involved in the estimation of the spatial model
was done as follows.
The Parametric Locally Stationary Convolution Model, see \ref{App_pLSM}, 
was used to generate  $n_{\rm repl}=33$ replicates of the true spectral functions $\sigma_\ell(x)$.
For each replicate of  $\boldsymbol\sigma^{(i)}(\cdot)$ (with $i=1,2,\dots,n_{\rm repl}$), we generated an ensemble
of $K$ pseudo-random realizations of $\xi\given\sigma$, from which
we computed vectors of sample band variances
${\bf d}^{(i)}(\cdot)$ at all analysis grid points following Eq.~(\ref{nlinInvProbl}). 
As a result, with $n_{\rm x}$ grid points,
we had the training sample of $n_{\rm train} = n_{\rm repl} \cdot n_{\rm x}$ pairs of vectors
${\bf d}^{(i)}(x)$ (inputs of the neural network) and $\boldsymbol\sigma^{(i)}(x)$ 
(targets to be predicted by the neural network).
The neural network was trained only once,
for the default setup.

\subsection {Results}
\label{sec_expm_anls_resu}

The accuracy of each  deterministic analysis was measured using the root-mean-square error 
(RMSE) with respect to the known truth.
The RMSE was computed by  averaging over the spatial grid and 100 independent analyses.
In this computation, each point of the spatial grid  had a weight proportional 
to the surface area of the respective grid cell.
In the figures below, we show the analysis performance score
\begin {equation}
\label{perf_score}
\frac{\rm RMSE - {\rm RMSE}_{\rm True B}}  {{\rm RMSE}_{\rm True B}},
\end {equation}
where ${\rm RMSE}_{\rm True B}$ is the RMSE of the optimal analysis.
We also show  90\% bootstrap confidence intervals (shaded).
In the bootstrap, we resampled the (independent by construction) spatial-grid-averaged mean-square errors.

We compared the five deterministic analyses (listed at the beginning of section \ref{sec_expm_anls})
in the default experimental setup and 
examined effects of varying the ensemble size $K$ and
the hyperparameters $\varkappa$  (controls the strength of the non-stationarity,  \ref{App_true_LSM_prm_flds})
and  $\mu_{\rm NSL}$  (controls the length scale of the non-stationarity,  \ref{App_true_LSM_pre}).

\begin{figure}[h]
\begin{center}
   { 
   \scalebox{0.4}{ \includegraphics{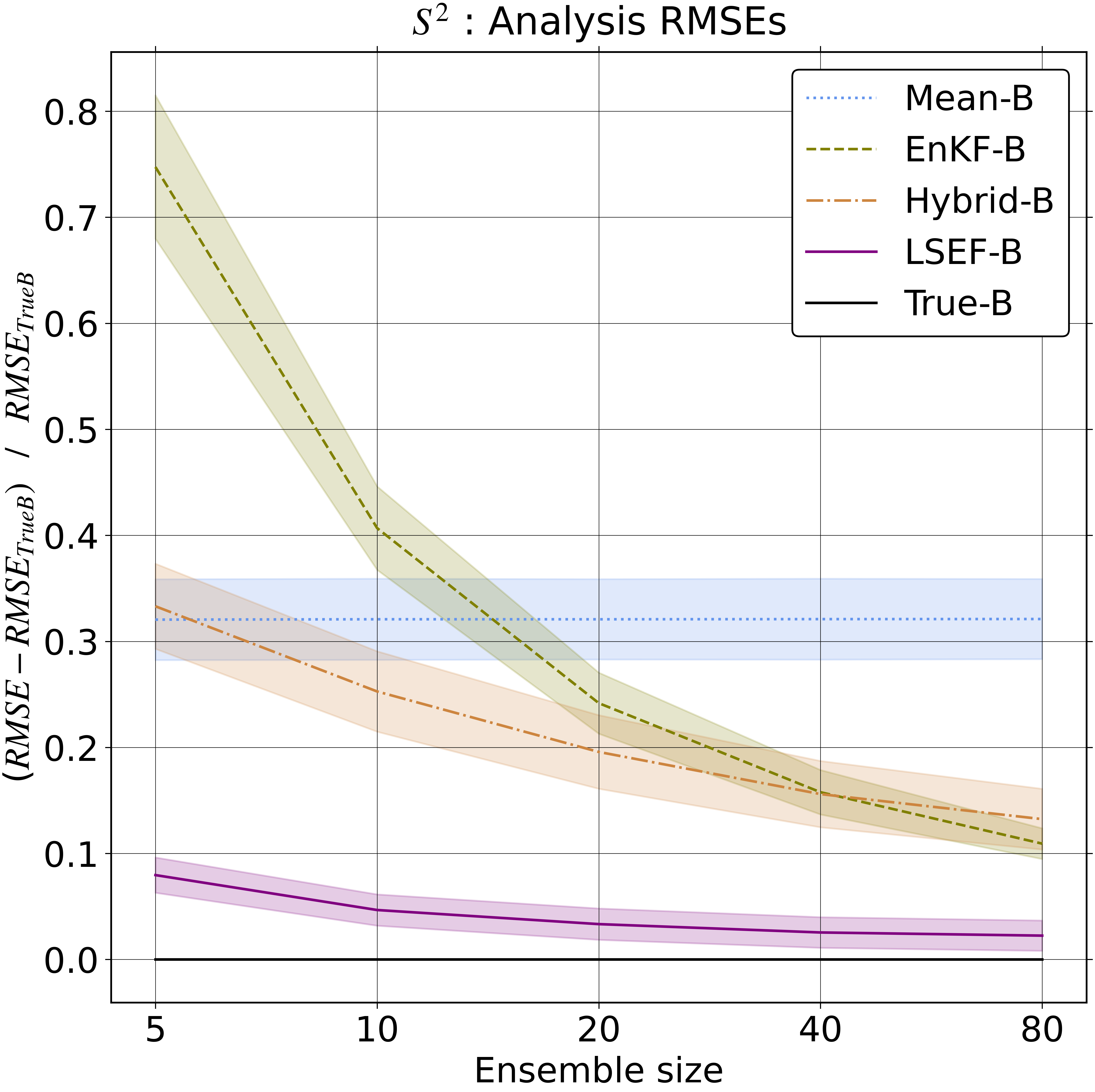}}
       }
\end{center}
  \caption{Analysis performance scores (the lower the better, =0 best possible),
  90\% bootstrap confidence bands (shaded).  Dependence on ensemble size.}
\label{Fig_an_esize}
\end{figure}

Figure~\ref{Fig_an_esize} shows dependencies of the RMSEs on the ensemble size.
It is seen that all ensemble-based analyses improved with the increasing $K$, as it should be.
The advantage of the LSEF-B analysis over the competitors was very significant.

\begin{figure}[h]
\begin{center}
   { 
   \scalebox{0.4}{ \includegraphics{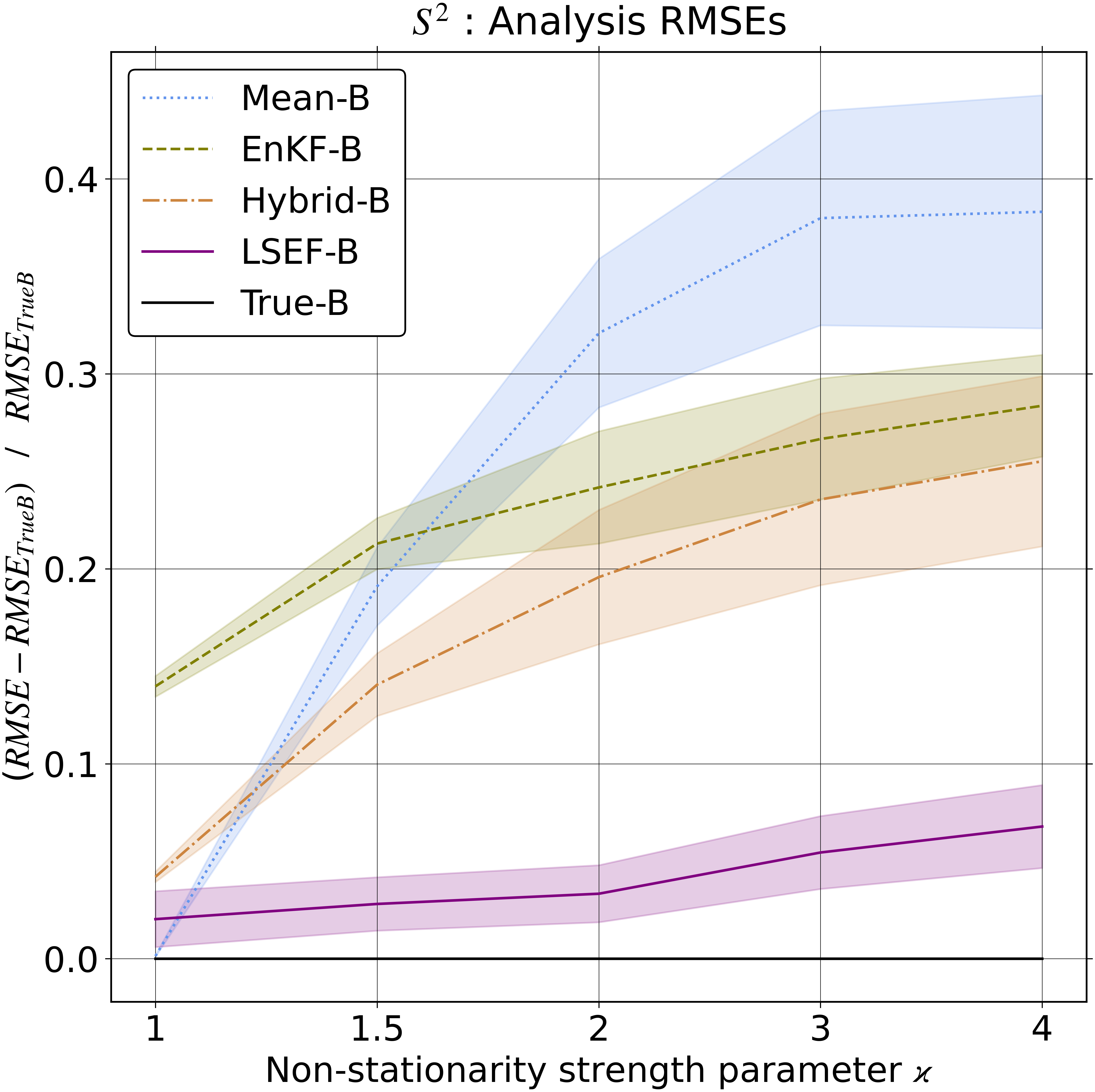}}
       }
\end{center}
  \caption{Analysis performance scores (the lower the better, =0 best possible), 
90\% bootstrap confidence bands (shaded). 
Dependence on the non-stationarity strength parameter $\varkappa$.}
\label{Fig_an_kappa}
\end{figure}

Figure~\ref{Fig_an_kappa} shows how the five analyses performed with the varying 
non-stationarity strength parameter $\varkappa$ (defined in \ref{App_true_LSM_prm_flds}).
The advantage of LSEF-B over EnKF-B was big and nearly uniform for all  $\varkappa >1$ on the plot.
It was Mean-B that suffered/benefited most from strong/weak non-stationarity, as expected.
In the stationary regime (\ie with $\varkappa =1$), Mean-B was almost as accurate as True-B
because ${\bf B}_{\rm mean}$ was a very accurate estimate of the constant (in this regime) ${\bf B}_{\rm true}$.
For this reason,  Mean-B  could not be 
rivaled by any ensemble-based analysis under stationarity.

\begin{figure}[h]
\begin{center}
   { 
   \scalebox{0.4}{ \includegraphics{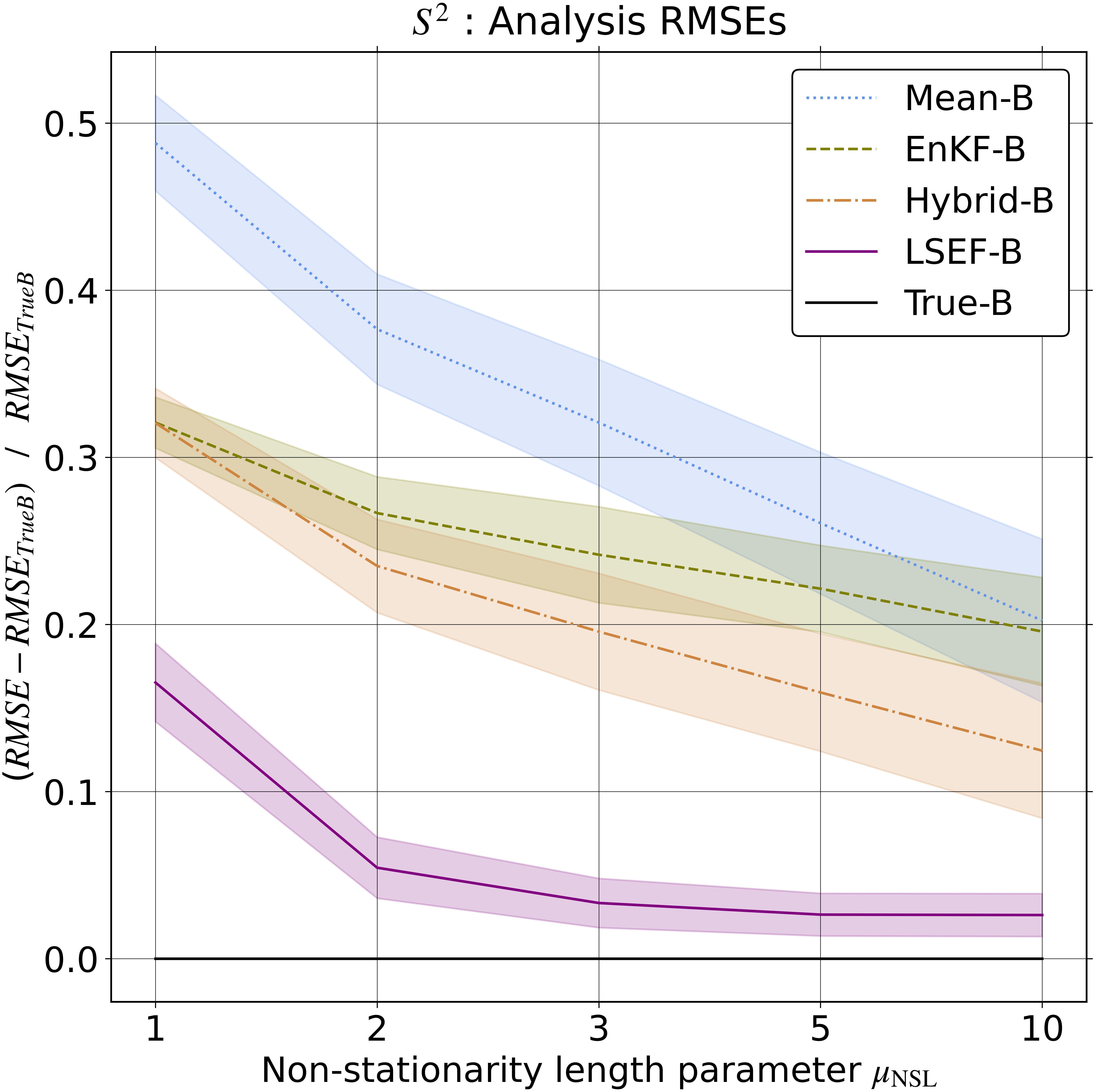}}
       }
\end{center}
  \caption{Analysis performance scores (the lower the better, =0 best possible),  
  90\% bootstrap confidence bands (shaded). 
Dependence on the non-stationarity length parameter $\mu_{\rm NSL}$.}
\label{Fig_an_NSL}
\end{figure}

Finally in this section, we study the impact of the non-stationarity length scale parameter
$\mu_{\rm NSL}$  (defined in \ref{App_true_LSM_pre}, see also section \ref{sec_expm_meth}) on the performance of LSEF. 
Figure~\ref{Fig_an_NSL} displays RMSEs of the five analyses as functions of $\mu_{\rm NSL}$. 
Again, LSEF-B was significantly better than the other sub-optimal analyses, and that was the case
even beyond the local stationarity assumption (section \ref{sec_LSM_locsta}), 
which requires that the non-stationarity length scale is much larger than the 
typical length scale of the field itself. 
Indeed, LSEF-B was substantially better than the competing techniques even when
$\mu_{\rm NSL}=1$ (see Fig.~\ref{Fig_an_NSL}), that is, when the non-stationarity length scale $\Lambda_{\rm NSL}$
was {\em equal} to the median length scale $L_{\rm med}$ (see \ref{App_true_LSM_pre}).

It is important to notice that in LSEF-B, there was no need to retrain the neural network for different setups. 
This is in contrast to EnKF-B, where we had to tune the localization length scale 
for each triple $K, \varkappa, \mu_{\rm NSL}$.
Similarly, ${\bf B}_{\rm mean}$ was computed for use in Mean-B and Hybrid-B for every pair  
$\varkappa, \mu_{\rm NSL}$. This suggests that the neural net-based LSEF-B was more robust than
the competing analyses.

The neural network was trained on the sample of the same size as we used to calculate
${\bf B}_{\rm mean}$, yet it allowed LSEF-B to exhibit much better results than Hybrid-B.
This suggests that the neural Bayes estimator
is capable of {\em selective} blending of sample covariances with time-mean covariances. This
is in contrast to Hybrid-B, which just mixes the two kinds of covariances with, effectively,
equal weights for all spatial scales.

It is worth noting that the neural-Bayes-based LSEF-B analysis significantly
outperformed an LSEF-B analysis based on the linear model estimator presented
in section \ref{sec_estm_consi} (not shown). This was the expected outcome because
the linear estimator (for which consistency is proven) is less accurate due to the linearity approximation.

\subsection{Effects of covariance regularization on spatial covariances}
\label{sec_eff_rglrz}

We reproduced the above experimental methodology (the  model of truth and the five analyses) 
on the circle and obtained very similar results (not shown).
The code was written by the other author in a different programming language (R \vs Python).

Here we take advantage of the simpler circular domain to
demonstrate how the regularization procedure proposed in this paper impacts  spatial covariances.
The spatial grid had 120 points and the ensemble size was 10.
 
\begin{figure}[h]
\begin{center}
   { 
   \scalebox{0.41}{ \includegraphics{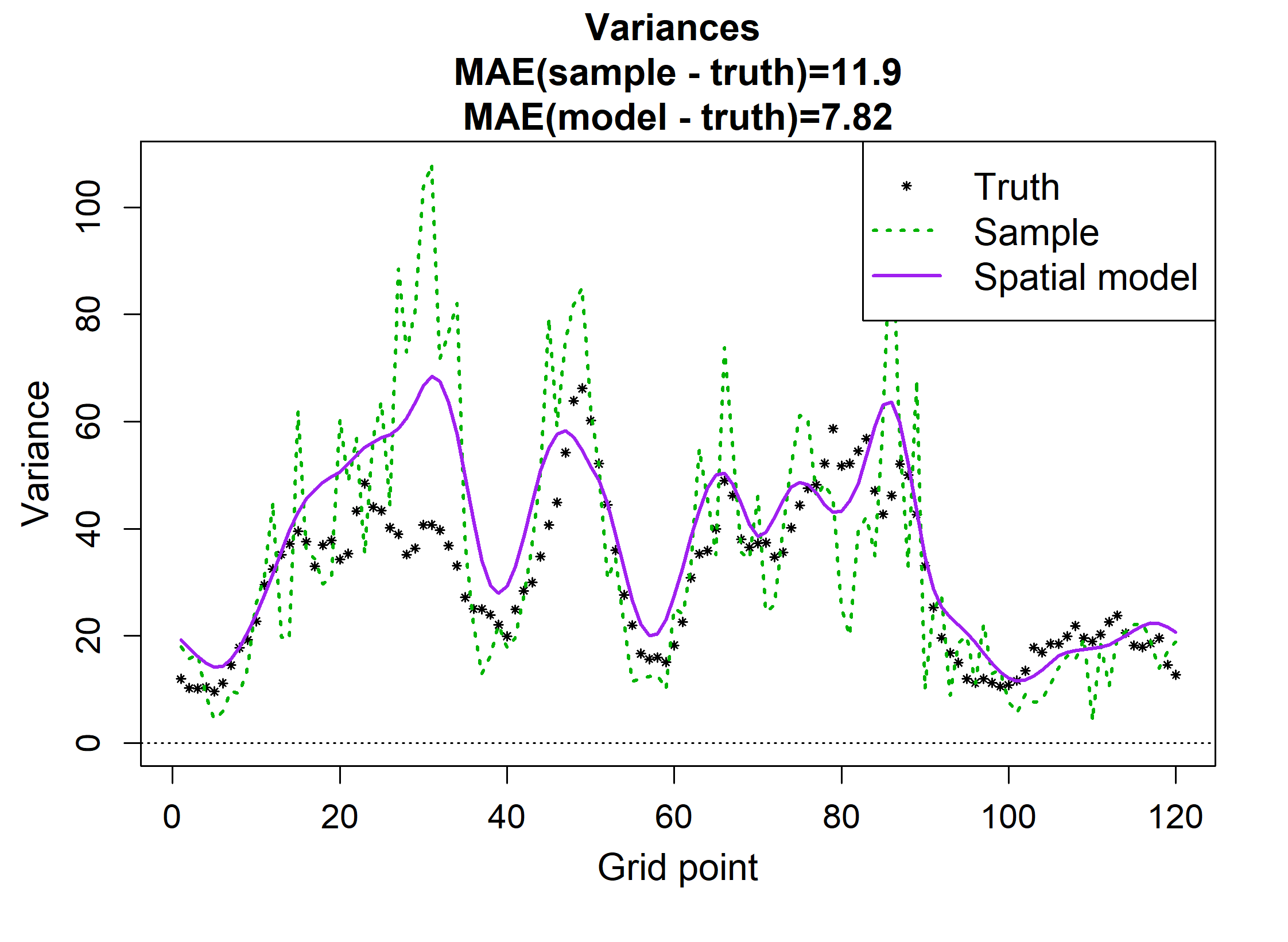}}
   \scalebox{0.41}{ \includegraphics{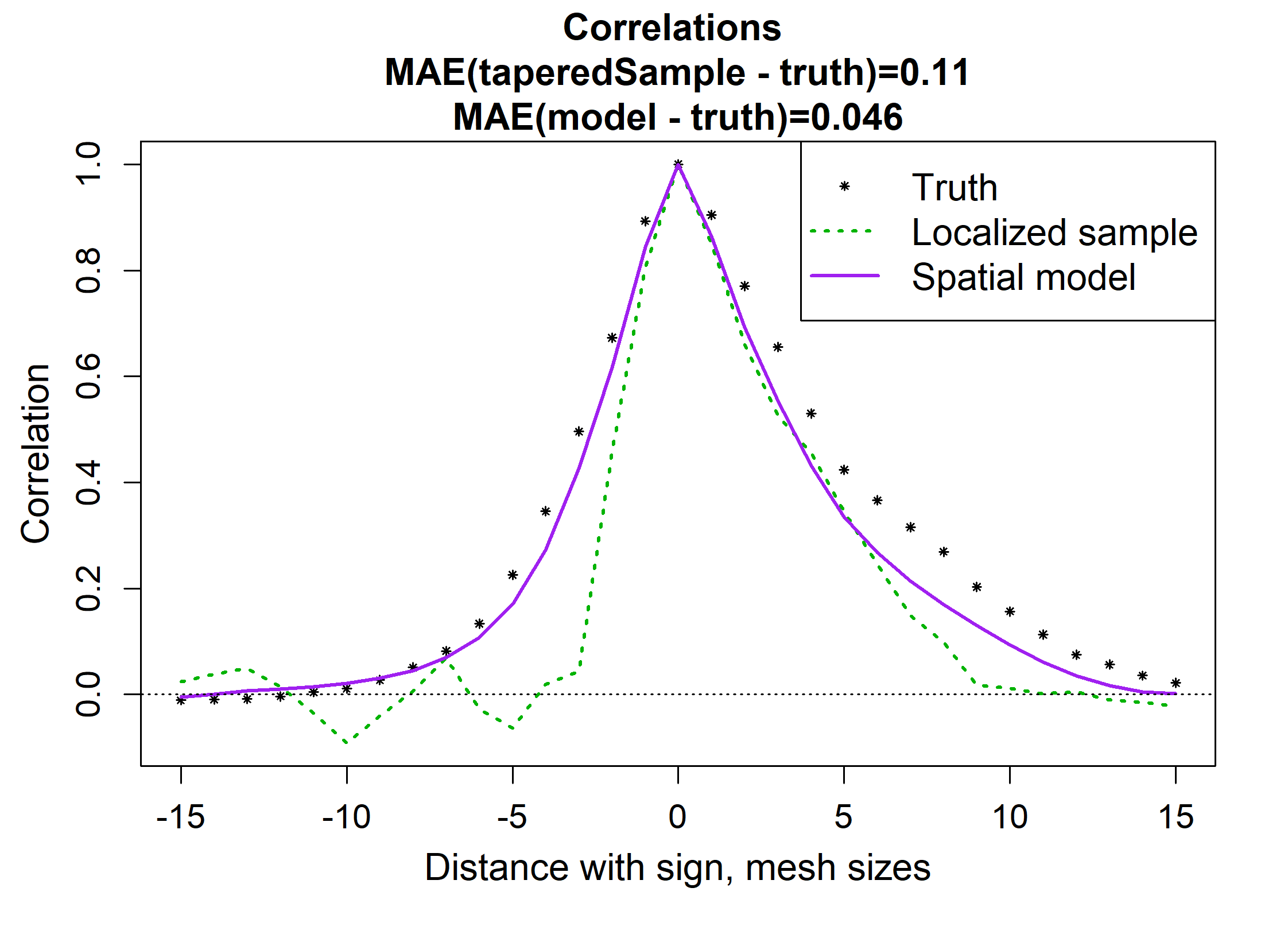}}
       }
\end{center}
  \caption{Comparison of typical true, localized sample (EnKF), and model (LSEF) covariances on the circle.
   Left: variances  at all grid points.
   Right: correlations with respect to a single grid point.}
\label{Fig_vars_corrs}
\end{figure}

Figure~\ref{Fig_vars_corrs} shows typical examples of how the true covariances (available
in the setting described above in this section) were
estimated from the ensemble by (i) the localized sample covariances 
(used in the baseline stochastic EnKF method) and (ii) the model covariances  
generated by the estimated Locally Stationary Convolution Model.
The left panel of Fig.~\ref{Fig_vars_corrs} shows the variances and the right panel the correlations\footnote{
The  realizations shown in Fig.~\ref{Fig_vars_corrs} were typical in the following sense. 
Their single-realization mean absolute errors  with respect to the truth 
were close (up to 10 percent difference) to the large-sample (of size 300) counterparts. 
For correlations, the mean absolute errors
were computed by averaging over distances up to 15 mesh sizes.}.

Figure~\ref{Fig_vars_corrs}(left) shows that the sample (EnKF's) variances are much noisier than the model variances.
This illustrates the fact that the EnKF's covariance localization cannot improve sample variances, 
which are quite uncertain for small ensemble sizes.
Less noise in the model (LSEF) variances is due to the regularizing effect of the constrained spatial model.
In particular, the local stationarity constraint implicitly
allows for the use of neighboring covariances in the estimation. 
The specific mechanism for that is provided by the LSEF's bandpass filters. 
 
The right panel of Fig.~\ref{Fig_vars_corrs} shows that, similar to the left panel, the
model correlations contain much less noise than their localized sample counterparts.
Another feature seen in this figure  is that, in contrast to model correlations, 
localized sample correlations suffer from a  negative bias in their length scale 
(see section \ref{sec_intro_literature}, item \ref{list_lcz}).
Furthermore, Fig.~\ref{Fig_vars_corrs}(right) illustrates that fact 
that the LSEF analysis produces almost noise-free large-distance background-error correlations 
without any ad-hoc localization (tapering) device.
This emphasizes the usefulness of the smooth-local-spectrum constraint 
(item \ref{list_constr_smoo_l} in section \ref{sec_constr}) 
and the ability of the neural Bayes estimator to ensure this constraint is met.
One can also notice in Fig.~\ref{Fig_vars_corrs}(right) 
that the regularizing constraint \ref{list_constr_loc_isotr} (section \ref{sec_constr}),
which requires the kernel to be locally isotropic, does not lead to isotropic correlations
(note the asymmetry in both the true and model correlations).
 
Averaged over 300 realizations of the truth and the ensemble, 
the mean absolute error in the model (LSEF) variances was about 1.5 times less
than the respective error in the sample variances. 
The mean absolute error in the model (LSEF)  spatial correlations was about {\em half} 
of the error in the localized sample correlations.
So, both   LSEF's variances and correlations were substantially more accurate than their localized sample counterparts.
This explains the significant advantage of the LSEF analysis over the 
EnKF analysis in the experiments presented above in this section.

\subsection{Conclusions on the experiments with static analyses}
\label{sec_static_conclu}

When the truth was generated by a (parametric) Locally Stationary Convolution Model, 
the proposed analysis technique performed very well compared with the competing techniques.
The question remains how the proposed approach performs in a more realistic situation
when background errors do not obey the Locally Stationary Convolution Model and the true local spectra 
are not just unavailable but they do not exist.
To answer this question we turn to sequential data assimilation.

\section{Numerical experiments with LSEF}
\label{sec_expm_LSEF}

Here we compare LSEF with the Kalman filter and the three sub-optimal filters whose analyses were
tested in section \ref{sec_expm_anls}.
In the sequential filtering experiments reported in this section, the spatial domain was the circle.

\subsection {Model of truth and forecast model}
\label{sec_LSEF_fcst_mdl}

To  test LSEF we took the Doubly Stochastic Advection-Diffusion-Decay Model \citep{DSADM}:
\begin {equation}
\label{DSADM}
{\bf x}_{t} = {\bf F}(\boldsymbol\theta_t) {\bf x}_{t-1} + \boldsymbol\Sigma_t \boldsymbol\eta_t,
\end {equation}
where
${\bf x}_{t}$ is the spatial field defined on a regular grid on the unit circle (a  vector),
 $t$ is the time index,
${\bf F}(\boldsymbol\theta_t)$ is the forecast model operator (a matrix),
$\boldsymbol\theta_t$ and  $\boldsymbol\Sigma_t$ are the spatial `parameter' fields (vectors)
defined below in Eqs.~(\ref{DSADM2}) and (\ref{DSADM3}), 
 $\boldsymbol\eta_t$ is the standard spatio-temporal white noise,
 and multiplication in the second term on the right-hand side is component-wise.
Double stochasticity means that both the forcing $\boldsymbol\eta_t$ and the parameters
$\boldsymbol\theta_t, \boldsymbol\Sigma_t$ are random.
$\boldsymbol\theta_t, \boldsymbol\Sigma_t$ are postulated to be stationary 
(in space and time) random fields satisfying the same model 
Eq.~(\ref{DSADM}) but with constant non-random coefficients 
$\theta_\theta, \theta_\Sigma, \Sigma_\theta, \Sigma_\Sigma$:
\begin {equation}
\label{DSADM2}
\boldsymbol\theta_{t} = {\bf F}(\theta_\theta) \boldsymbol\theta_{t-1} + 
    \Sigma_\theta \boldsymbol\eta_t',
\end {equation}
\begin {equation}
\label{DSADM3}
\boldsymbol\Sigma_{t} = {\bf F}(\theta_\Sigma) \boldsymbol\Sigma_{t-1} + 
    \Sigma_\Sigma \boldsymbol\eta_t',
\end {equation}
where  $\boldsymbol\eta_t', \boldsymbol\eta_t'$ are  
independent standard spatio-temporal  white noises.
Variability in $\boldsymbol\theta_t, \boldsymbol\Sigma_t$ leads to non-stationarity 
(in space and time) of the conditional distribution
${\bf x} \,| (\boldsymbol\theta, \boldsymbol\Sigma)$, \ie when
 $\boldsymbol\theta_t, \boldsymbol\Sigma_t$ are simulated and then fixed in the simulation of ${\bf x}_t$.

The reason for using this model was its ability to generate non-stationary 
random fields   ${\bf x}_t$ without
sacrificing linearity. Linearity of the forecast 
model allowed us to compare the filters with the unbeatable benchmark, 
the exact Kalman filter.
This is normally not possible with nonlinear forecast models.

In the filters, the forecast model operator was ${\bf F}(\boldsymbol\theta_t)$ and the forcing was zero.
First, we generated the parameter fields $\boldsymbol\theta_t, \boldsymbol\Sigma_t$ and computed 
the model error covariance matrices 
${\bf Q}_{t} = \boldsymbol\Sigma_t \boldsymbol\Sigma_t^\textsf{T}$ 
(made available  to the Kalman filter only).
Then,  $\boldsymbol\theta_t, \boldsymbol\Sigma_t$ 
were fixed and we computed the truth ${\bf x}_{t}$ (as a solution to Eq.~(\ref{DSADM}))
and observations ${\bf x}_{t}^{\rm obs}$ (as the noise-contaminated truth at some grid points and 
every analysis time instant $t$). 
After that, having fixed the truth and the observations, we ran the filters.

As a result, we were able to set up and examine the Kalman Filter (KF), LSEF, and the three competing 
ensemble filters: EnKF, the Mean-B filter, and the Hybrid-B filter.

\subsection{Filtering setup}
\label{sec_flt_setup}

The Doubly Stochastic Advection-Diffusion-Decay Model was defined on a 120-point regular grid on the circle
(so that  $\ell_{\rm max}=60$).
Otherwise, its setup coincided with that reported in \citet{DSADM}.

The ensemble size was between 5 and 160 with the default 10.
The number of bandpass filters and the setup of the neural network were the same as 
in the above experiments on the sphere.
5000 consecutive assimilation cycles were used to validate the filters.
${\bf B}_{\rm mean}$ was computed by averaging Kalman filter's ${\bf B}$ over 100,000 cycles
and, additionally, by averaging in space.
The latter was done by averaging entries of  ${\bf B}$ over its main diagonal (getting the mean variance)
and each cyclic super-diagonal (getting the whole set of mean covariances). 
`Synthetic' observations were located at every 10th model grid point.
The observation error variance was specified to roughly match the mean Kalman-filter 
forecast error variance (this resembles the relationship between observation and forecast errors in meteorological
data assimilation, our application area).

In the EnKF  and Hybrid-B filters,  multiplicative covariance inflation  
\citep[e.g.][]{Houtekamer2016} was used and tuned to optimize their performance. 
Ensemble perturbations were multiplied by a factor of 1.02. 
There was no covariance inflation in LSEF (as we found that it deteriorated its performance, not shown).

The performance of the LSEF analysis reported in section \ref{sec_expm_anls} was so
impressive that we did no tuning there (other than  choosing the neural network's hyperparameters). 
We recall that the reason for that was the setting in which (i) the prior uncertainty was, by construction, governed
by the Locally Stationary Convolution Model and (2) the hyperprior of the model parameters was also available
to train the neural network. 
Here, as  neither of these two conditions was met, some tuning was done. 
First, we found it beneficial  to increase the number
of bandpass filters from 6 to 8. Second, we specified ${q_{\rm shape}}=2$ and 
the same width $\Delta_{{\cal H}_j}=\Delta_{{\cal H}}$ of the spectral
transfer function  (see Eq.~(\ref{Hj})) for all bandpass filters. 
Third,  we retrained  the neural network for each ensemble size.
Fourth, we  increased  $\Delta_{{\cal H}}$ for larger ensemble sizes:
$\Delta_{{\cal H}}=5$ for $n_{\rm e} \le 40$ and $\Delta_{{\cal H}}=10$ for $n_{\rm e} > 40$
(following the finding in \ref{App_consi} that it is worth broadening the spectral
transfer functions of the bandpass filters with the growing ensemble size, see also section \ref{sec_estm_consi}).
Fifth, we specified the (hyper)prior sample of the spectral functions $\sigma_\ell(x)$ 
in a different, ad-hoc way compared to section \ref{sec_expm_anls}, see the next subsection.

\subsection{Training sample}
\label{sec_expm_asml_train}

In the course of sequential filtering with LSEF, 
the forecast error field does not have to obey the Locally Stationary Convolution Model,
its imposition in the LSEF analysis is an approximation. Therefore we had no access
to true local spectra here (as they do not exist in this setting) and thus could not immediately apply the 
technique described in section \ref{sec_expm_anls_train} to train the neural network.

To build the training sample, we initially tried to manually set up a 
circular analogue of the Parametric Locally Stationary Convolution Model 
(described in  \ref{App_pLSM}) in such a way that the spatial non-stationarity of 
the pseudo-random fields generated by it resembled those of the forecast-error fields we
obtained in the course of Kalman filtering.
In that way, we were able to generate the training sample following section \ref{sec_expm_anls_train}.
However, the filtering results were mediocre (not shown).
Another option we tested was to fit, having the EnKF forecast ensemble at every assimilation cycle, 
the  Locally Stationary Convolution Model.
The fitted model provided us with the estimated local spectra, which were then used in the training sample
as the `truth'.
The respective sample band variances were computed
from a new ensemble generated by the fitted model.
This approach  did yield an improvement, but the following technique
was simpler and somewhat more successful,
so it was used in the experiments presented below.

During an EnKF run, at each assimilation cycle $t=1,2,\dots, 10^5$, 
we performed {\em spatial averaging} of the forecast-ensemble sample covariance matrices ${\bf S}_t$
(in the same way the  spatial averaging was done in the computation of ${\bf B}_{\rm mean}$, see 
section \ref{sec_flt_setup}). The resulting 
spatially averaged covariance functions $b_t(x)$ (where $x=0,1,\dots, n_{\rm x}/2$ labels the distance
measured in mesh sizes) were stored.
Then, offline, for each  covariance function $b_t(x)$ we computed its spectrum (Fourier transform) 
${\bf f}_t =(f_t(0), f_t(1), \dots, f_t(\ell_{\rm max}))$.
Knowing ${\bf f}_t$, we computed, for each $t$, an ensemble of pseudo-random
realizations of the respective (stationary) random process, subjected them to the spatial 
bandpass filters (section \ref{sec_estm_bandpass_flt}), 
and finally computed the vectors ${\bf d}_t$ of sample band variances (section \ref{sec_estm_two_stage}).
The resulting training sample consisted of  100,000  pairs $({\bf d}_t, \sqrt{{\bf f}_t})$.
The above spatial averaging of the EnKF sample covariances reduced the diversity in the training sample.
To compensate for that effect, we set up the Doubly Stochastic Advection-Diffusion-Decay Model 
in the strongly non-stationary regime \citep[][Table 3]{DSADM} when running EnKF to
generate the training sample.

\subsection{Results}
\label{sec_expm_flt_resu}

The performance of each of the five filters was measured using the RMSE (with respect to the truth)
of their control forecasts ${\bf x}^{\rm f}_{t}$.
The RMSE was computed by  averaging over the spatial grid, over 5000 assimilation cycles, and
over 10 replicates of the  parameter fields 
 $\boldsymbol\theta_t, \boldsymbol\Sigma_t$ of the  model of truth (outlined in section \ref{sec_LSEF_fcst_mdl}).
As in section \ref{sec_expm_anls_resu}, we show here the analysis performance score,  Eq.~(\ref{perf_score}), 
where ${\rm RMSE}_{\rm True B}$ is the RMSE of Kalman filter forecasts.

We compared  the five filters for different ensemble sizes  and 
in four regimes of non-stationarity of the true spatio-temporal field.
The regimes are labeled by numbers from 0 (stationarity) to 3 (strong  non-stationarity). 
For a more detailed description of the non-stationarity regimes, see \citet{DSADM}.
The default non-stationarity regime was 2.

\begin{figure}[h]
	\begin{center}
		{ 
			\scalebox{0.52}{ \includegraphics{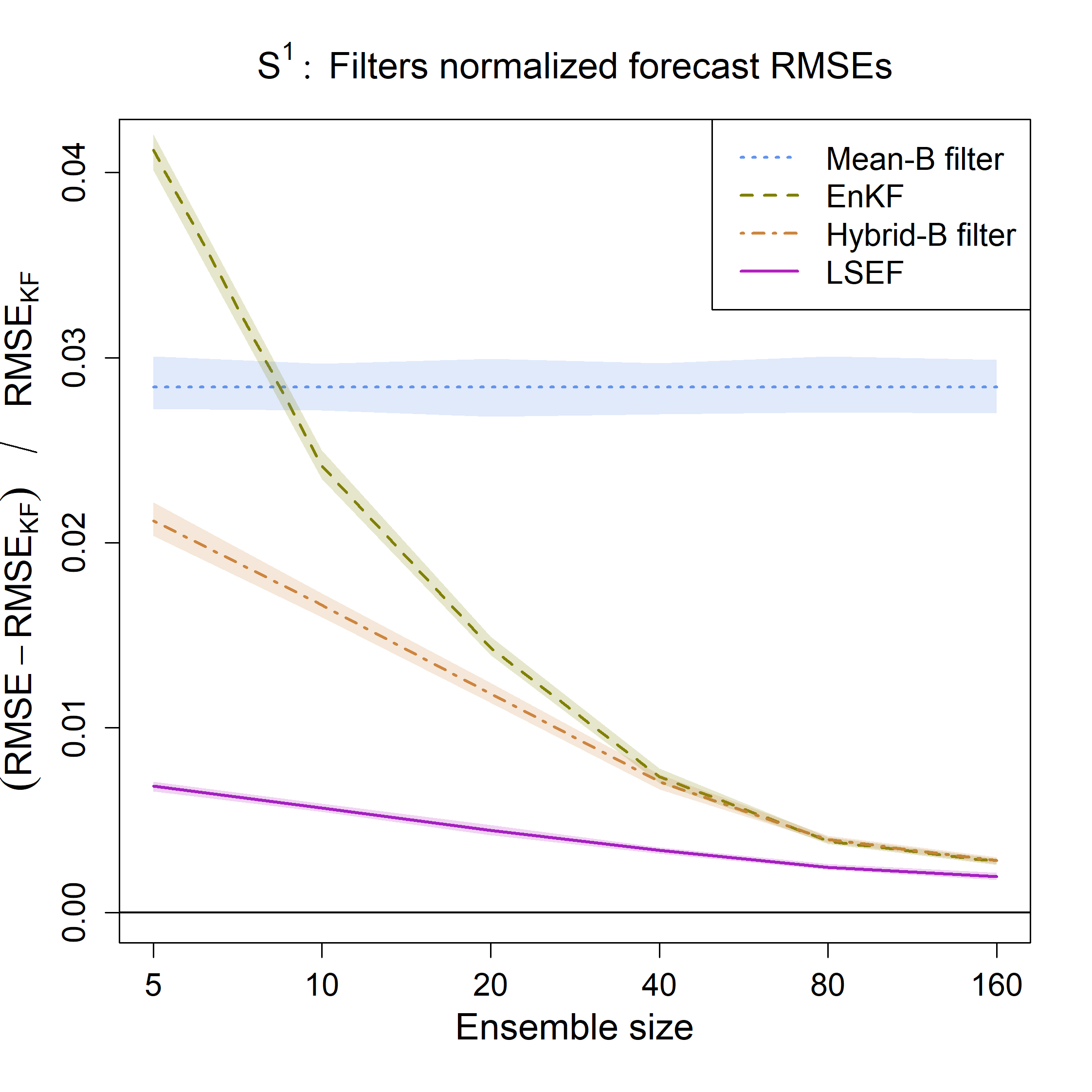}}
		}
	\end{center}
	\caption{Filters  performance scores (the lower the better, =0 best possible),  
90\% bootstrap confidence bands (shaded). 
Dependence on ensemble size.}
	\label{Fig_LSEF_ne}
\end{figure}

Figure~\ref{Fig_LSEF_ne} shows
that LSEF was significantly better than the competing EnKF, Mean-B, and Hybrid-B filters.
Comparing Fig.~\ref{Fig_LSEF_ne} with 
Fig.~\ref{Fig_an_esize} demonstrates that the great advantage of LSEF
observed when the truth was generated by the Parametric 
Locally Stationary Convolution Model (Fig.~\ref{Fig_an_esize})
was more modest in a more realistic situation here (Fig.~\ref{Fig_LSEF_ne}).
Nevertheless, the LSEF outperformed the three competing approximate filters
even for very large ensemble sizes.
Note that for a system with 120 degrees of freedom, 
the ensemble of size 160 is indeed very large, a situation we never have
in high-dimensional systems.
For  $K \le 40$, the superiority of LSEF was substantial.

\begin{figure}[h]
\begin{center}
   { 
   \scalebox{0.52}{ \includegraphics{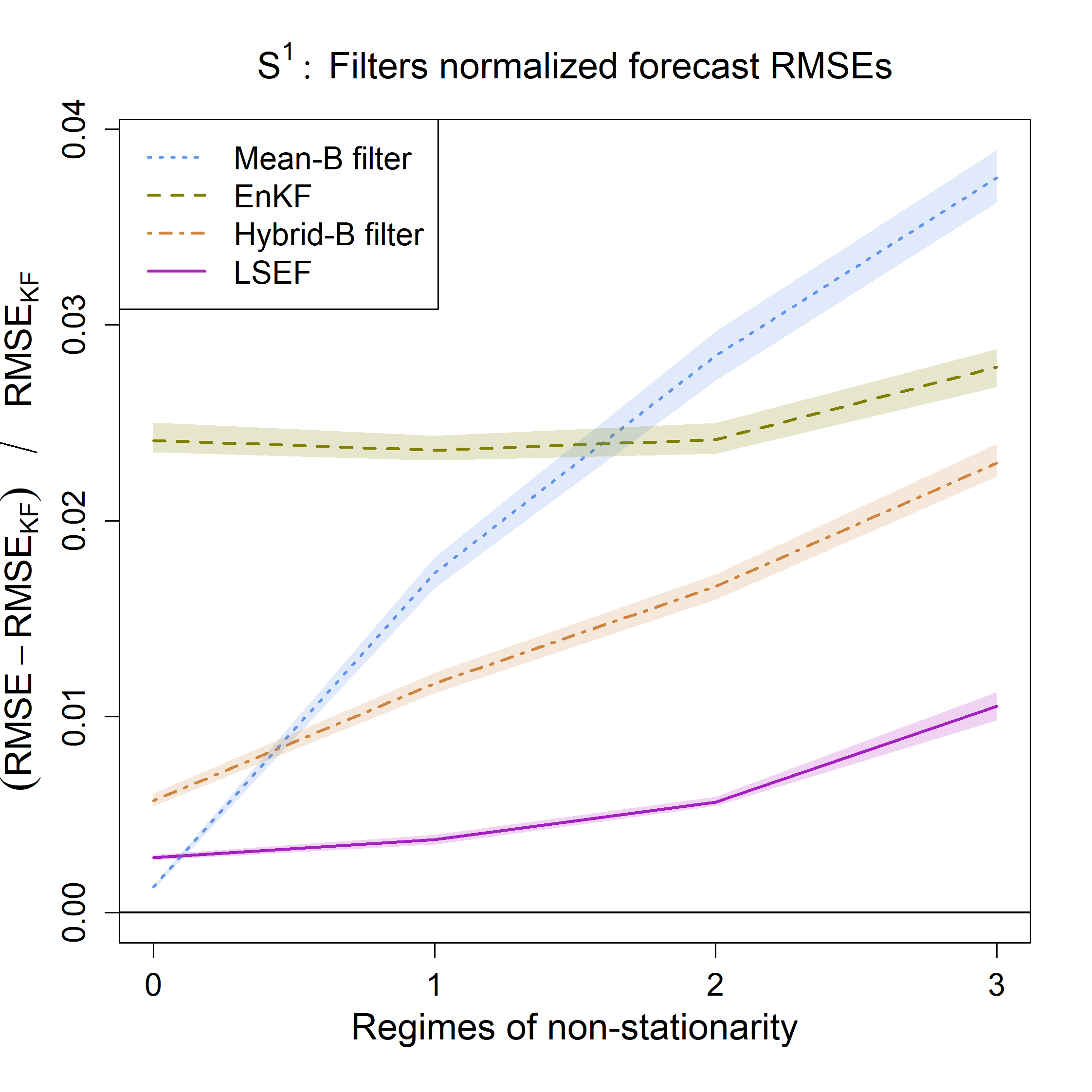}}
       }
\end{center}
  \caption{Filters performance scores (the lower the better, =0 best possible),  
90\% bootstrap confidence bands (shaded). 
Dependence on the non-stationarity regime (0 is stationary, 1 is weakly non-stationary, 
2 is default, 3 is highly non-stationary).}
\label{Fig_LSEF_regimes}
\end{figure}

Figure~\ref{Fig_LSEF_regimes} shows how the five filters performed under different 
non-stationarity regimes. The superiority of LSEF was nearly uniform except for the stationary 
regime 0.
Like in section \ref{sec_expm_anls_resu} (see Fig.~\ref{Fig_an_NSL} at $\mu_{\rm NSL}=1$), 
the advantage of LSEF remained substantial even under strong non-stationarity (regime 3 on the 
 x-axis in Fig.~\ref{Fig_LSEF_regimes}).

It is also worth mentioning that with each ensemble-based filter examined in our experiments,
the spread (standard deviation) of the forecast ensemble did match the forecast RMSE pretty well.
The difference was within 3\% of the RMSE.

\section{Discussion and conclusions}
\label{sec_conclu}

This research aims to enhance the ensemble Kalman filter by introducing a spatial model at the analysis 
(observation update) step.
The idea is to abandon the use of sample covariances, which are
notoriously noisy in high-dimensional problems.
Instead, we propose to build a non-parametric, non-stationary spatial model, constrain it
to reduce its number of degrees of freedom, and 
estimate it directly from  the prior ensemble. The estimated model is then used
to specify prior covariances in the analysis.

The  model is formulated (on the sphere and the circle, but can be extended to other domains)
for the underlying forecast-error random field rather than for its spatial covariances,
which guarantees that the implied covariance matrices are positive definite.
The model is non-stationary in space, 
which can be essential for real-world (\eg geoscientific) applications, and
non-parametric, which has the potential of producing 
more diverse  shapes of spatial correlations compared to those produced by  parametric models.
The spatial model is formulated as a Gaussian process convolution model.
The spatially varying convolution kernel $u(x,\rho)$  is {\em constrained} to reduce the 
effective number of degrees of freedom and regularize the estimation task.

One fundamental regularizing constraint is 
local stationarity, hence the name, the Locally Stationary Convolution Model. 
Local stationarity means that the random field (process) in question
can be approximated by a stationary process locally in space.
Local stationarity takes place whenever the typical length scale of the process is 
significantly less
than a {\em non-stationarity length scale}, the scale on which the local structure of the process varies in space.
In other words, whenever the  convolution kernel $u(x,\rho)$  changes 
slowly with location ($x$) compared with its change as a function of distance ($\rho$).
The stationary process-convolution model is a special case of our non-stationary model.

Another fundamental constraint imposed on the spatial model is formulated in spectral space 
for {\em spectral functions} $\sigma_\ell(x)$ defined to be the spectral transform of  $u(x,\rho)$ with respect to $\rho$.
We require $\sigma_\ell(x)$ 
(or the {\em local spectra} $f_\ell(x) = \sigma^2_\ell(x)$) 
to be {\em smooth} functions of the wavenumber $\ell$. 
This constraint means that there is no need for high spectral resolution in estimating the spectral functions $\sigma_\ell(x)$.
The relatively low spectral resolution allows for a higher resolution in physical space,
thus permitting to capture spatial non-stationarity in the model estimator.
Besides, smooth local spectra imply rapidly decaying spatial correlations.
As a result, spurious long distance correlations are suppressed
(without additional ad-hoc devices like localization or tapering), which
is important in data assimilation applications.

The constraints are part of a (hyper)prior model for the model parameters. 
The list of constraints can be modified or extended.
For example, we may know the typical degree of {\em roughness/smoothness} of the analyzed field
(determined by the behavior of  spatial correlations at small distances). 
With our spatial model, this knowledge can be 
encoded in a constraint on the rate of decay of the spectral functions $\sigma_\ell(x)$
at large wavenumbers.

In general, using  a spatial model  to regularize the analysis step in a high-dimensional ensemble filter
is a  way to incorporate problem-specific spatial information into the analysis.
This information may be available or can be acquired by a dedicated effort.

Aiming at high-dimensional applications, we built a computationally efficient model
estimator.
Relying on the local stationarity and smooth-local-spectra constraints, 
the estimator employs spatial bandpass filters to extract aggregated local spectra from the input 
ensemble of random field realizations. 
The aggregated local spectra are then disaggregated (deconvolved) from sample variances of the filtered fields.
With a linear disaggregation technique, we prove that the bandpass filters can be specified in such a way that
the resulting model estimator is consistent.
A more accurate nonlinear disaggregation method based on the use of a neural network is developed.

Due to the regularizing constraints, the estimator produces much less noise in the resulting
spatial covariances compared with the traditional localized (tapered) sample covariances.
However, it does so at the expense of introducing a systematic distortion in the 
covariances/spectra---because the imposed constraints need not be respected in real-world spatial processes.
The main hypothesis underlying this research is that there are filtering problems in which
the reduction of noise in spatial covariances (\ie the reduced random uncertainty) 
has a greater positive impact on the filter's accuracy than the negative effect of the
induced systematic uncertainty in the filter's prior distribution.
Our numerical experiments suggest that this hypothesis is plausible.

We postulated that in the course of sequential ensemble filtering,
the prior uncertainty is governed by the Locally Stationary Convolution Model
estimated online from the prior ensemble and a hyperprior distribution.
This allowed us to optimize the use of both the ensemble and the forecast-error `climatology'.
As a result, we obtained a 
new sequential filter termed the Locally Stationary convolutional Ensemble Filter (LSEF).
Testing the filter with a non-stationary spatio-temporal model of truth yielded encouraging results.
It is worth noting that ad-hoc regularization devices like 
localization and covariance inflation, which are, normally, used
in EnKF and hybrid filters
were absent in LSEF. This can be important in practice since
ad-hoc devices require extensive (and expensive) tuning.
Another interesting result is that the advantage of LSEF was observed not only
under weak non-stationarity, as intended, 
but in strongly non-stationary regimes as well.

Our spatial-model-based covariance regularization technique was tested 
with linear observation models on the sphere and
 linear forecast and observation models on the circle.
We believe that linearity of these tests is not a significant restriction, expecting that 
the technique will be useful whenever an 
EnKF-like filter is useful, that is, with non-linear models as well.

Using a neural Bayes estimator to recover local spectra 
from the outputs of the spatial bandpass filters 
proved to be successful even in the situation when true hyperprior
distribution was {\em not} available to train the neural net
(in sequential filtering).
In case when the  true hyperprior was  available
(in our experiments with static analyses on the sphere), 
the advantage of the neural-Bayes-estimator based analysis was dramatic.
Our results also suggest that placing a hyperprior on the spatial model's parameters leads to better extraction of
`climatological' information from the training sample than the mixing of ensemble prior covariances with
the time-mean prior covariance matrix
(the popular technique reviewed in section \ref{sec_intro_literature}, see item \ref{list_hybr}).
One reason for this is that the hyperprior distribution is more informative than just the mean
covariance matrix. Another reason is that the expressivity of a neural network is much greater than that of 
a linear combination with scalar weights.

Applying the developed  regularization technique to real-world high-dimensional problems
will require a 
multigrid/multi-scale version of the algorithm. Work on it is underway. 
The key idea is to approximate a  given convolution operator using sparse
operators defined on a hierarchy of spatial grids. 
The computational cost will be drastically reduced by representing 
large-distance spatial interactions on coarse grids.
 

Adapting our model to fields with strong non-stationarities that occur at fixed known locations
(like ocean coastlines) can also be subject of future research.
Further extensions of the spatial model may include directional anisotropy, and, possibly, non-Gaussianity.
In applications to non-global domains, spectral bandpass filtering can be replaced with
spatial wavelet  filtering.

Last but not least, a promising avenue for further research is to extend 
the approach proposed in this paper to deal with imperfect prior ensembles.
This can be especially important if the ensemble is generated by a fast data-driven forecast model.
Even though such an ensemble can be large, it can suffer from probability biases.
The role of the spatial model will then be to  correct for those biases.


\section*{Acknowledgments}
\label{sec_ack}

We would like to thank Dmitry Gayfulin, who 
contributed to the development of the regularization technique
at an early stage of this study.
Constructive criticism of two anonymous reviewers of the first (rejected) version of the manuscript is highly appreciated.
Valuable comments of two anonymous reviewers and the editor, Alfred Stein, 
of the second version of the manuscript allowed us to substantially improve and clarify the manuscript.
The study has been conducted
under Task  1.1.1  of the Scientific Research and Technology 
Plan of the Russian Federal Service for
Hydrometeorology and Environmental Monitoring.


\appendix

\section{Approximate sequential filtering}
\label{App_lin_seq_flt}

We review the Kalman filter and sub-optimal non-ensemble and
ensemble filters that are used in our numerical experiments.
The sub-optimal (approximate) filters are reviewed in the linear setting
sufficient for our purposes in this paper.
For more details and a broader perspective, see, \eg \citet{Asch,Houtekamer2016}.

\subsection{Linear filtering setup}
\label{App_lin_flt_setup}

Here, we add a few notational and technical details to what is presented in section \ref{sec_seq_flt} above.

By  $n_{\rm x}$ we denote the dimensionality of the system, \ie the length of its state vector ${\bf x}_{t}$.
By $n_t^{\rm obs}$ we denote the number of observations available at the time instant $t$.
The  model-error covariance matrix is $\Ex \boldsymbol\eta_t\boldsymbol\eta_t^\TT = {\bf Q}_t$.
The  observation-error covariance matrix is $\Ex \boldsymbol\varepsilon_t\boldsymbol\varepsilon_t^\TT = {\bf R}_t$.

The time-discrete linear filtering aims at finding {\em mean-square optimal}  estimates and their 
uncertainties.
A mean-square optimal estimate is a linear combination of all available data (observations)
such that the estimation error is orthogonal (with respect to the covariance inner product) 
to the available data.

\subsection{Forecast step}
\label{App_fcst_bckg}

In the optimal (Kalman) filter, assume  we have the {\em optimal analysis} vector  ${\bf x}^{\rm a}_{t-1}$, that is,
the point estimate of  ${\bf x}_{t-1}$ whose error is orthogonal to the observations  
${\bf x}^{\rm obs}_{t-1}, {\bf x}^{\rm obs}_{t-2}, \dots$,
along with the analysis error covariance metrix ${\bf A}_{t-1}$.
Then, the {\em optimal forecast} vector 
(the point estimate of  ${\bf x}_{t}$ from the same observations)
is  simply the deterministic (control) forecast
\begin {equation}
\label{fc}
{\bf x}^{\rm f}_{t} = {\bf F}_t {\bf x}^{\rm a}_{t-1}
\end {equation}
provided that the model error $\boldsymbol\eta_t$ is uncorrelated with
the observation error $\boldsymbol\varepsilon_{t'}$ for any $t'<t$. 
It is straightforward to write down the error covariance matrix of the 
deterministic forecast:
${\bf B}_{t} = {\bf F}_t {\bf A}_{t-1} {\bf F}_t^\TT + {\bf Q}_t$.
So, the Kalman filter propagates in time the optimal point estimate of the truth and
its error covariance matrix.

In a sub-optimal (ensemble or non-ensemble) filter, the forecast Eq.~(\ref{fc}) is no longer optimal 
but it is unbiased if so is ${\bf x}^{\rm a}_{t-1}$ 
(subtract Eq.~(\ref{evolu}) from Eq.~(\ref{fc})).
For this reason, we use Eq.~(\ref{fc}) to compute the control forecast in 
the experiments presented in this paper (see also  Remark in  \ref{App_anls_ens}).

Instead of propagating the error covariance matrix, an ensemble filter
propagates an {\em ensemble} that represents the uncertainty in ${\bf x}^{\rm f}_t$:
\begin {equation}
\label{fe}
{\bf x}^{{\rm fe}(k)}_{t} = {\bf F}_t {\bf x}^{{\rm ae}(k)}_{t-1} + \boldsymbol\eta_t^{(k)},
\end {equation}
where $k=1,2,\dots,K$ labels the ensemble member, $K$ is the ensemble size,
the analysis-ensemble members ${\bf x}^{{\rm ae}(k)}_{t-1}$ are defined below in \ref{App_anls_ens}, and
the simulated model error 
$\boldsymbol\eta_t^{(k)}$ is drawn from a probability distribution that, ideally
(and in the experiments in this study), coincides with 
 the distribution of the true model error $\boldsymbol\eta_t$.

\subsection {Non-ensemble analysis}
\label{App_anls_nonens}

In a non-ensemble analysis scheme, we are given, first, the forecast 
(analysis background) ${\bf x}^{\rm f}$, see Eq.~(\ref{fc}).
(Till the end of this subsection, the time subscripts are omitted since all vectors correspond to the same time instant.)
Second, we have current observations  ${\bf x}^{\rm obs}$.
Third, a static proxy ${\bf B}_{\rm static}$ to the true background-error covariance matrix 
${\bf B}$ is available.
The  point estimate of the truth (often called the deterministic or control analysis) ${\bf x}^{\rm a}$ 
is sought as a linear combination of ${\bf x}^{\rm f}$
and ${\bf x}^{\rm obs}$.

If (i) the true background-error covariance matrix ${\bf B}$ is known,
(ii) ${\bf x}^{\rm f}$ is an optimal estimate of the truth by itself
(meaning that no affine transformation of the forecast is better in the mean-square sense),
and (iii) $\boldsymbol\varepsilon$ is probabilistically independent of 
the forecast error  $\boldsymbol\xi = {\bf x}^{\rm f} - {\bf x}$ 
(this is the case if $\boldsymbol\varepsilon$ is independent of all past observation and model errors),
then the {\em optimal} (Kalman filter) analysis scheme can be employed:
\begin {equation}
\label{xa} 
{\bf x}^{\rm a} = {\bf x}^{\rm f} +  {\bf K} ({\bf x}^{\rm obs} - {\bf H} {\bf x}^{\rm f}),
\end {equation}
where 
\begin {equation}
\label{K} 
{\bf K} =  {\bf B }{\bf H}^\TT ({\bf H}{\bf B }{\bf H}^\TT + {\bf R})^{-1} \equiv
({\bf B}^{-1} + {\bf H}^\TT {\bf R}^{-1}{\bf H})^{-1} {\bf H}^\TT {\bf R}^{-1}
\end {equation}
is the (Kalman) gain matrix. The optimal analysis error covariance matrix is
${\bf A} =  ({\bf B}^{-1} + {\bf H}^\TT {\bf R}^{-1}{\bf H})^{-1} \equiv ({\bf I - KH)B}$.

In a {\em sub-optimal} non-ensemble analysis scheme, only a proxy
${\bf B}_{\rm static}$ to the true background-error covariance matrix 
is available. 
In this situation, the common practice  
is  to  plug in ${\bf B}_{\rm static}$
instead of ${\bf B}$ into Eq.~(\ref{K}) and use the resulting gain matrix to compute the analysis vector
using Eq.~(\ref{xa}). 
This approach is taken in so-called variational data assimilation  \citep[e.g.][]{Asch}.

\subsection {Ensemble analysis}
\label{App_anls_ens}

The ensemble analysis scheme has on input the control forecast  ${\bf x}^{\rm f}$ and the 
forecast ensemble  $\{ {\bf x}^{{\rm fe}(k)} \}_{k=1}^{K}$, see Eq.~(\ref{fe}).
The static covariance matrix ${\bf B}_{\rm static}$ can also be available.
The goal is to compute the deterministic analysis vector ${\bf x}^{\rm a}$ 
and the {\em analysis ensemble}
 $\{ {\bf x}^{{\rm ae}(k)} \}_{k=1}^{K}$  aimed to characterize the error in ${\bf x}^{\rm a}$.

{\em Remark}. With nonlinear forecast models, the  deterministic forecast ${\bf x}^{\rm f}$ 
is normally replaced with the ensemble mean forecast $\overline{{\bf x}^{\rm fe}}$.
However, with the linear forecast model, linear observations, and zero-mean forecast and
observation errors,
the {\em deterministic forecast} is unbiased (see below in this subsection and 
 \ref{App_fcst_bckg}). Therefore it is a better
proxy to the truth than the ensemble mean. 
Indeed,  
$\overline{{\bf x}^{\rm fe}} - {\bf x} \equiv (\overline{{\bf x}^{\rm fe}} - {\bf x}^{\rm f}) + 
({\bf x}^{\rm f} - {\bf x}) = -
\frac{1}{K} \sum \boldsymbol\xi^{(k)} + \boldsymbol\xi$,
where $\boldsymbol\xi^{(k)} = {\bf x}^{\rm fe} - {\bf x}^{{\rm fe}(k)}$
are the pseudo-random {\em ensemble perturbations} and $\boldsymbol\xi$ is the error 
in the  control forecast.
If $\Ex\boldsymbol\xi=0$ and $\boldsymbol\xi^{(k)}$
are independent of  $\boldsymbol\xi$, then
the mean square error of the ensemble mean is seen to be greater than that of 
the control forecast.
For this reason, we  used  ${\bf x}^{\rm f}$ rather than $\overline{{\bf x}^{\rm fe}}$
to compute the deterministic analysis in our experiments.

The ensemble analysis  computes the gain matrix $\widehat{\bf K}$ using  Eq.~(\ref{K}) in which
${\bf B}$ is replaced with its estimate, $\widehat{\bf B}$,
computed from the ensemble ${\bf x}^{{\rm fe}(k)}$ and, possibly,  from the
static covariances ${\bf B}_{\rm static}$.
Recent-past estimated covariances can also be utilized, see 
references in   section \ref{sec_intro_literature}, item \ref{list_intro_recent}.
The Ensemble Kalman Filter (EnKF) explicitly or implicitly 
relies on the sample covariance matrix ${\bf S}$ to specify the prior covariances. 
In the so-called stochastic variant of EnKF
(which is used in our experiments), $\widehat{\bf B}$ equals ${\bf B}_{\rm EnKF}$, which is 
${\bf S}$ element-wise multiplied 
by a so-called localization matrix. 
Hybrid ensemble-variational data assimilation schemes use a  convex combination 
of ${\bf B}_{\rm static}$  and ${\bf B}_{\rm EnKF}$ to specify $\widehat{\bf B}$.

The resulting gain matrix  $\widehat{\bf K}$ is  used to compute the
deterministic analysis  ${\bf x}^{\rm a}$ following Eq.~(\ref{xa}).
Since  $\widehat{\bf K}$ is sub-optimal, ${\bf x}^{\rm a}$ is not optimal either,
but it is unbiased if so is the control forecast.
Indeed, writing 
$\widehat{\bf K} = \widehat{\bf K}(\boldsymbol\xi^{(1)}, \boldsymbol\xi^{(2)},\dots,\boldsymbol\xi^{(K)})$
to indicate its dependence on the forecast ensemble perturbations $\boldsymbol\xi^{(k)}$,
substituting 
$\widehat{\bf K}$ into Eq.~(\ref{xa}), and
using Eq.~(\ref{obs}), we obtain the analysis error
\begin {equation}
\label{xa_ee} 
{\bf x}^{\rm a} - {\bf x} = \boldsymbol\xi +  
  \widehat{\bf K}(\boldsymbol\xi^{(1)},\boldsymbol\xi^{(2)},\dots,\boldsymbol\xi^{(K)})
                  \cdot (\boldsymbol\varepsilon - {\bf H} \boldsymbol\xi),
\end {equation}
which has zero expectation   because $\boldsymbol\xi$  and
$\boldsymbol\varepsilon$ are zero-mean
random vectors  independent of $\boldsymbol\xi^{(k)}$.

As for the analysis ensemble, we confine our attention to the so-called perturbed-observations 
approach to generate its members:
\begin {equation}
\label{xae_ens} 
{\bf x}^{{\rm ae}(k)} = {\bf x}^{{\rm fe}(k)} + 
  \widehat{\bf K} \cdot ({\bf x}^{\rm obs} + \boldsymbol\varepsilon^{(k)} - {\bf H}{\bf x}^{{\rm fe}(k)}),
\end {equation}
where $\boldsymbol\varepsilon^{(k)}$ 
are independent pseudo-random draws from the true observation-error distribution.

\section{Non-negative definiteness of the non-stationary model}
\label{App_pos_def}

By definition, a real-valued symmetric kernel, $f(x,x')$, 
is non-negative definite (and thus is a valid covariance function
of a  random field) if 
for any set of spatial points $x_1, x_2,\dots x_N$ and any real 
numbers $c_1, c_2, \dots, c_N$, it holds that
\begin {equation}
\label{sum}
\sum_{p=1}^N \sum_{q=1}^N c_p c_q f(x_p,x_q) \ge 0.
\end {equation}
The  spatial covariance function of our non-stationary spatial convolution model is
given by Eq.~(\ref{BxxS2}).
Applying the addition theorem of spherical harmonics, we rewrite Eq.~(\ref{BxxS2})  as
\begin {equation}
\label{BxxS2_}
B(x,x') =  \sum^{\ell_{\rm max}}_{\ell=0} \sigma_\ell(x) \, \sigma_\ell(x')
                          \sum_{m=-\ell}^\ell Y_{\ell m} (x) \, Y_{\ell m}^* (x'),
\end {equation}
where  $\sigma_\ell(\cdot)$ are real-valued.
Non-negative definiteness of $B(x,x')$ immediately follows from
\begin {equation}
\label{sum2}
\sum_{p=1}^N \sum_{q=1}^N c_p c_q B(x_p,x_q) = \sum^{\ell_{\rm max}}_{\ell=0} \sum_{m=-\ell}^\ell
  \left| \sum_{p=1}^N  c_p \sigma_\ell(x_p)  Y_{\ell m} (x_p) \right|^2
\ge 0.
\end {equation}
%

\section{The spatial model on the circle}
\label{App_LSM_S1}

On $\S^1$, the model Eq.~(\ref{osc2}) specializes to 
\begin {equation}
\label{xiu_S1}
\xi(x) = \int_{\S^1} u(x, y-x) \,\alpha(y) \,\d y,
\end {equation}
where the real-valued convolution kernel $u(x, \rho)$ is an even function of the second argument.
Let us employ the truncated Fourier series expansion of the kernel 
in the form
\begin {equation}
\label{u_spe_S1}
u(x, \rho) =  \frac{1}{\sqrt{2\pi}}   \sum_{\ell=-\ell_{\rm max}}^{\ell_{\rm max}}  \sigma_\ell(x) \e^{\i \ell \rho}.
\end {equation}
Here $\sigma_\ell(x)$ are real-valued functions such that $\sigma_{-\ell}(x)=\sigma_\ell(x)$.
With the positive-definite convolution kernel (section \ref{sec_pos_def_kernel}), $\sigma_\ell(x) > 0$.
The band-limited Gaussian white noise $\alpha$ in Eq.~(\ref{xiu_S1}) has the spectral expansion
\begin {equation}
\label{alpha_S1}
\alpha(y) =  \frac{1}{\sqrt{2\pi}}  \sum_{\ell=-\ell_{\rm max}}^{\ell_{\rm max}}  \widetilde\alpha_{\ell}  \e^{\i \ell y},
\end {equation}
 $\widetilde\alpha_\ell$ are mutually uncorrelated zero-mean and unit-variance  random variables.
$\widetilde\alpha_0$ is real-valued and the others are complex-valued circularly symmetric random variables.

Substituting Eqs.~(\ref{u_spe_S1}) and (\ref{alpha_S1}) into Eq.~(\ref{xiu_S1}) 
yields the spectral version of the spatial model 
on the circle:
\begin {equation}
\label{LSM_S1}
\xi (x) =  \sum_{\ell=-\ell_{\rm max}}^{\ell_{\rm max}}  \sigma_\ell(x)  
                          \,\widetilde\alpha_{l} \e^{\i \ell x},
\end {equation}
for which the  local spectrum is defined as $f_\ell(x) = \sigma_\ell^2(x)$.
From Eq.~(\ref{LSM_S1}) it follows that
the respective covariance function on $\S^1$ is
\begin {equation}
\label{BxxS1}
B(x,x') = \Ex \xi (x) \,\xi (x') =  \sum_{\ell=-\ell_{\rm max}}^{\ell_{\rm max}}
                          \,\sigma_\ell(x) \, \sigma_\ell(x') \e^{\i \ell (x'-x)}.
\end {equation}
%

\section{The most localized kernel in the stationary case}
\label{App_kernel_lcz}

Here we consider the stationary process convolution model.
We show that its convolution kernel  is not unique given the model's 
covariance function (or, equivalently, the spectrum). Uniqueness can be achieved by 
requiring that the kernel be most spatially localized.

Let us define a stationary (isotropic) process on the sphere through its spectral representation:
\begin {equation}
\label{xi_statio_spe}
\xi_{\rm statio} (x) = \sum^{\ell_{\rm max}}_{\ell=0} \sum^\ell_{m=-\ell} \,\widetilde\xi_{\ell m} \, Y_{\ell m} (x) \equiv
                       \sum^{\ell_{\rm max}}_{\ell=0} \sum^\ell_{m=-\ell} \,\sigma_\ell  \,\widetilde\alpha_{\ell m} \, Y_{\ell m} (x),
\end {equation}
where it is straightforward to see that  
$\Ex \widetilde\xi_{\ell m} \widetilde\xi_{\ell' m'}^* = \sigma_\ell^2 \, \delta_\ell^{\ell'} \delta_{m}^{m'}$.
The (real-valued) spectrum of the process is $f_\ell = \sigma_\ell^2$.
The process  $\xi_{\rm statio}(x)$ can be modeled as 
the convolution of the kernel  
\begin {equation}
\label{urho}
u(\rho) =  \sum^{\ell_{\rm max}}_{\ell=0} \frac{2\ell +1}{4\pi} \sigma_\ell P_\ell(\cos\rho)
\end {equation}
with the white noise $\alpha$ (see Eq.~(\ref{alpha})):
\begin {equation}
\label{xi_statio}
\xi_{\rm statio}(x) = \int_{\S^2} u(\rho(x,y)) \,\alpha(y) \,\d y.
\end {equation}
The kernel $u(\rho)$ is unique given the set of $\sigma_\ell$ but it is not unique 
given the spectrum $f_\ell = \sigma_\ell^2$ of the process $\xi_{\rm statio}(x)$ --- just because
$\sigma_\ell$ is defined by $f_\ell$ up to sign.
To isolate a unique kernel, we require it to be {\em most spatially localized} in the sense
that it has a minimal width. We define the width as a kind of macro-scale from 
\begin {equation}
\label{R_S2}
R^2_{\rm u} = \frac {\int_{\S^2} u^2(\rho(x,y)) \,\d y} {(\max |u(\rho)|)^2}.
\end {equation}
Here, the numerator is equal to $ \sum \frac{2\ell +1}{4\pi} f_\ell$ and thus is fixed given the spectrum $f_\ell$.
So, $R_{\rm u}$ is minimized when  $\max |u(\rho)|$ is maximal among all kernels with the fixed $|\sigma_\ell|$.
That is, we seek to maximize $|u(\rho)|$ over both $\rho$ and the signs of $\sigma_\ell$.
From Eq.~(\ref{urho}), we have
\begin {equation}
\label{u_FL}
|u(\rho)| =  \left|  \sum^{\ell_{\rm max}}_{\ell=0} \frac{2\ell +1}{4\pi}  \sigma_\ell P_\ell(\cos \rho) \right| \le
             \sum^{\ell_{\rm max}}_{\ell=0} \frac{2\ell +1}{4\pi}  |\sigma_\ell|
\end {equation}
because $|P_\ell(t)| \le 1$ for $-1\le t \le 1$
\citep[][section 7.21]{Szego}.
Since $|P_\ell(t)| = 1$ if and only if $t = \pm 1$,
the upper bound (\ie the maximum of  $|u(\rho)|$) indicated in Eq.~(\ref{u_FL}) is 
reached if all $|P_\ell(\cos \rho)|=1$ (which implies that $\rho=0$ or $\rho=\pi$)
and all products  $\sigma_\ell P_\ell(\cos \rho)$ have the same sign.
This happens in three cases so  there are three solutions to the optimization problem $R_{\rm u} \to \min$.

The first solution is  $\sigma_\ell > 0$ for all $\ell$. The corresponding kernel
$u_1(\rho)$ is positive definite and its modulus is maximized at $\rho=0$.

The second solution is  $\sigma_\ell \le 0$ for all $\ell$. The corresponding kernel $u_2(\rho)$
is non-positive definite and
  its modulus is maximized  at $\rho=0$ so that $u_2(\rho) = -u_1(\rho)$.

The third solution is $\sigma_\ell = (-1)^\ell |\sigma_\ell|$. The corresponding kernel $u_3(\rho)$
is non-definite,
its modulus is maximized at $\rho=\pi$,  and $u_3(\rho) = u_1(\pi-\rho)$
(this follows from the identity $P_\ell(-t)=(-1)^\ell P_\ell(t)$).

As the above three kernels that minimize the length scale $R_{\rm u}$  have the same shape, we select
the first solution: the {\em positive definite kernel} $u(\rho)$.

On the circle, similar arguments lead to virtually the same conclusion:
the most localized kernel is a positive definite function $u(x)$ or its negated/translated version
$\pm u(x-h)$ (for any $h \in \S^1$, the proof is omitted).

\section{Identifiability of the model}
\label{App_identif}

Here we examine the  process convolution model with the locally isotropic real-valued 
convolution kernel $u(x,\rho(x,y))$. The kernel $u(x,\rho)$ is a positive definite function of $\rho$ for any $x$ 
(equivalently, the spectral functions $\sigma_\ell(x)$ are positive for any $x$).
We prove that if the non-stationary covariance function $B(x,x')$ is produced by this model, 
then the model is unique, 
\ie  $\sigma_\ell(x)$ are uniquely determined by $B(x,x')$.
For simplicity, we consider the circular case.

Let us start with expanding  $\sigma_\ell(x)$ into the truncated Fourier series
\begin {equation}
\label{sigma_lq}
\sigma_\ell(x) = \sum_{q=-Q}^{Q}    
                          \,\widetilde\sigma_{\ell q} \e^{\i  q x},
\end {equation}
where $Q$ is the half-bandwidth of the processes $\sigma_\ell(x)$ as functions of location $x$.
Note  that since  $\sigma_{-\ell}(x) = \sigma_\ell(x)$
(see  \ref{App_LSM_S1}), we have 
$\widetilde\sigma_{-\ell, q} = \widetilde\sigma_{\ell q}$.
Since  $\sigma_\ell(x)$ is real-valued, we have 
$\widetilde\sigma_{\ell, -q} = \widetilde\sigma_{\ell q}^*$.

We  assume that the bandwidth $[-Q,Q]$ is tight in the sense that  $\widetilde\sigma_{\ell Q} \ne 0$ for all $\ell$.

\begin{figure}
\begin{center}
   {
   \scalebox{0.15}{ \includegraphics{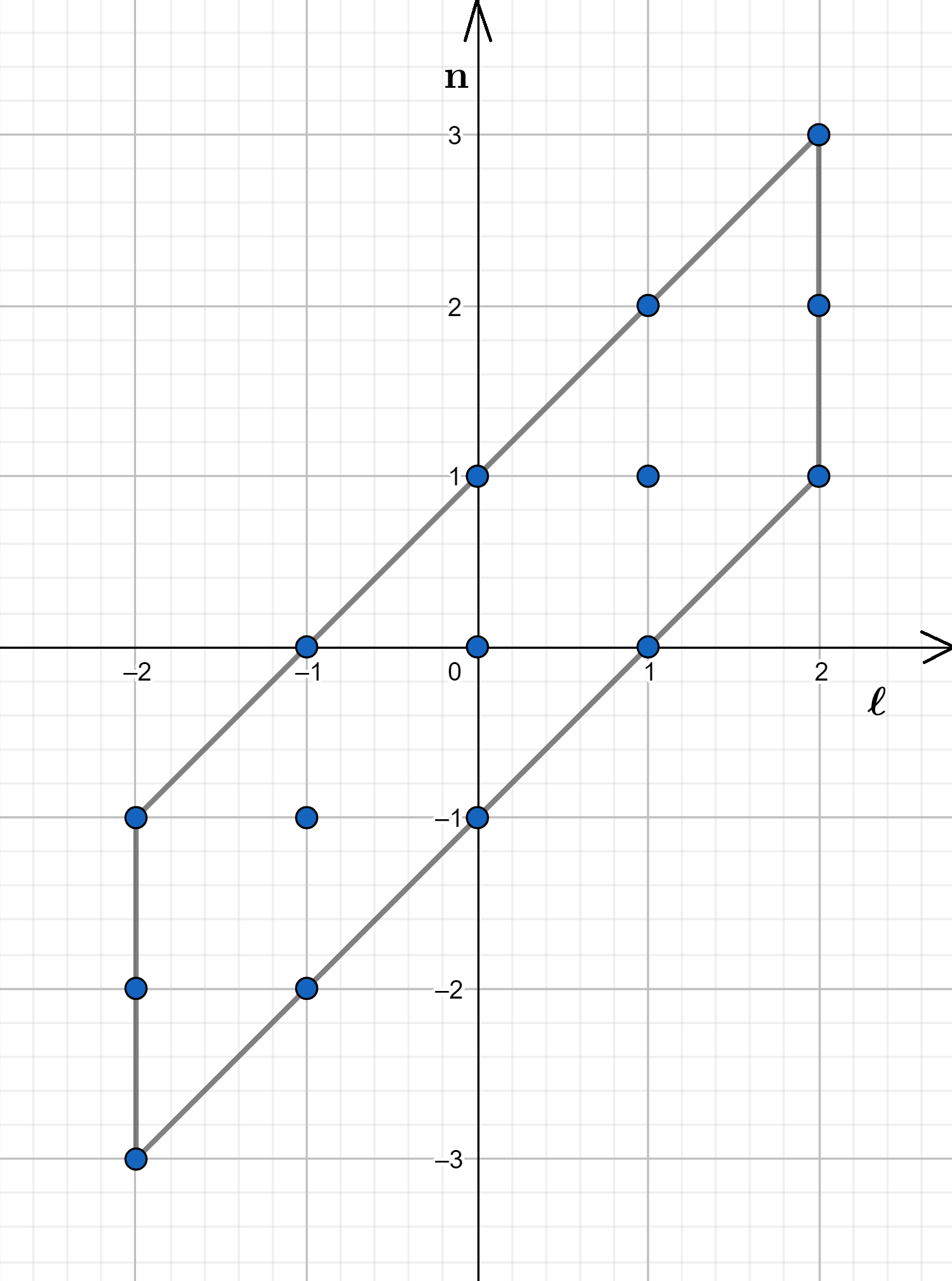}}
       }
\end{center}
  \caption{Circular case. Summation area for $L=2$ and $Q=1$}
\label{Fig_App_summ}
\end{figure}

Let us substitute the expansion Eq.~(\ref{sigma_lq}) into Eq.~(\ref{LSM_S1}) and change the summation
variable $q$ to $n=\ell+q$\footnote{
In this Appendix (in contrast to the rest of the paper), 
we denote $\ell_{\rm max}$ by $L$ to simplify the notation.
}:
\begin {equation}
\label{xi_ln}
\xi (x) =  \sum_{\ell=-L}^{L} \widetilde\alpha_{\ell} \e^{\i \ell x} 
           \sum_{q=-Q}^{Q} \,\widetilde\sigma_{\ell q} \e^{\i  q x}  =
   \sum_{n=-N}^{N}   \widetilde\xi_n \e^{\i n x},
\end {equation}
where $N=L+Q$ and
\begin {equation}
\label{xi_n}
\widetilde\xi_n =   \sum_{\substack{\ell=n-Q \\ |\ell| \le L}}^{n+Q} 
   \widetilde\alpha_{\ell} \,\widetilde\sigma_{\ell,n-\ell}.
\end {equation}
The summation area in Eq.~(\ref{xi_n}) for all possible $n$ and $\bar n$ is 
shown in Fig.~\ref{Fig_App_summ} for $L=2$ and $Q=1$.

Since $\xi(x)$ is uniquely represented by the set of its spectral coefficients $\{\widetilde\xi_n \}_{n=-N}^N$,
the covariance function  $B(x,x')$ is uniquely represented by the set of the spectral covariances
$\widetilde B_{n \bar n} = \Ex \widetilde\xi_n \widetilde\xi_{\bar n}^*$.

Our goal is to prove that all $\widetilde\sigma_{\ell q}$
are uniquely determined by the set of  $\widetilde B_{n \bar n}$
(and thus by the covariance function  $B(x,x')$).

We start with  $n=N \equiv L +Q$, for which the sum in Eq.~(\ref{xi_n}) reduces to the single term,
$\widetilde\xi_N = \widetilde\alpha_{L} \,\widetilde\sigma_{L Q}$
(the upper right corner of the summation area in Fig.~\ref{Fig_App_summ}).
This means that the covariance  $\widetilde B_{Nn}=\Ex\widetilde\xi_N\,\widetilde\xi_n^*$ for any
$n=L+q \in [L-Q, \, L+Q]$ will also contain just one term,
\begin {equation}
\label{BNn}
\widetilde B_{Nn} \equiv \widetilde B_{L+Q, L+q} = 
  \Ex \widetilde \xi_{L+Q} \, \widetilde \xi_{\ell+q}^* = 
     \widetilde\sigma_{L Q} \, \widetilde\sigma^*_{Lq},
\end {equation}
 so that we can uniquely restore $\widetilde\sigma_{Lq}$ for all $q$.
To realize this, take $q=-Q$ in Eq.~(\ref{BNn})
(the lower right corner  of the summation area in Fig.~\ref{Fig_App_summ}).
We obtain
$\widetilde B_{N,L-Q} = \widetilde\sigma_{L Q} \,\widetilde\sigma_{L,-Q}^*$.
Recalling that 
 $\widetilde\sigma_{\ell, -q} = \widetilde\sigma_{\ell q}^*$, we have
$\widetilde B_{N,L-Q}=\widetilde\sigma_{L Q}^2$, which implies that $\widetilde\sigma_{L Q}$ 
is uniquely determined by $\widetilde B_{N,L-Q}$, up to  sign. 
By the tight bandwidth assumption, 
$\widetilde\sigma_{L Q} \ne 0$, therefore, we  recover  
$\widetilde\sigma_{Lq}$ for all $|q|<Q$ from Eq.~(\ref{BNn}) 
(the right edge of the 
summation parallelogram in Fig.~\ref{Fig_App_summ}), again, up to  sign. 
The ambiguity in the sign disappears when it comes to $\widetilde\sigma_{L0}$, which is 
positive because $\sigma_\ell(x) > 0$ due to the positive-definiteness constraint 
(see item \ref{list_constr_pos_def} in section \ref{sec_constr} and  \ref{App_LSM_S1}).
Thus, $\widetilde\sigma_{Lq}$ are uniquely recovered  for all $q$ from the model covariances. 
 
Then, we consider $\widetilde\xi_{N-1}$ and realize that it, again, has just one term,
$\widetilde\alpha_{L-1} \,\widetilde\sigma_{L-1,L-1+Q}$, besides the term that contains 
$\widetilde\sigma_{L,L+Q-1}$, which has already been recovered. This allows us to repeat the above process
and  recover $\widetilde\sigma_{L-1,L-1+q}$ for all $q\in[-Q,Q]$. 
And so on, we uniquely recover all non-zero $\widetilde\sigma_{\ell q}$,
and therefore all  spectral functions $\sigma_\ell(x)$, from the set of spectral covariances $\widetilde B_{n \bar n}$.
This completes the proof that the non-stationary spatial convolution model on the circle is unique
provided that the convolution kernel $u(x,\rho)$ is real-valued, is an even function of $\rho$,
and the half-bandwidth $Q$ of  $\sigma_\ell(x)$ is such that  $\widetilde\sigma_{\ell Q} \ne 0$ for any $\ell$.
On $\S^2$, the same reasoning is valid (not shown).

\section{Bandpass filtering of a locally stationary signal}
\label{App_accu_est}

Here we consider spatial bandpass filtering of a random process governed by the Locally Stationary  Convolution Model.
We consider the filtered process 
in the locally stationary regime with large $\Lambda$ or, formally, in the  limit $\Lambda \to\infty$, with
the mean process spectrum $\Ex\sigma_\ell^2$ (see section \ref{sec_LSM_locsta}) being kept constant.
We examine the approximation errors involved in Eqs.~(\ref{phi1}) and Eq.~(\ref{vj})
and show that these errors  are small whenever $\Lambda$ is much greater than
the the length scale of the filter's impulse response function 
(formally, they vanish in the above limit).
For simplicity of presentation, we consider the circular case.

Let the bandpass filter ${\cal H}$ have the spectral transfer function 
\begin {equation}
\label{H_kappa}
H(\ell) = \kappa \left(\frac{\ell-\ell^{\rm c}}{T} \right),
\end {equation}
where $\kappa(\cdot)$ is a smooth real-valued 
bell-shaped function of continuous argument such that
$\kappa(0)=\max|\kappa(\cdot)|=1$ and $\max|\kappa'(\cdot)| = 1$,
$\ell^{\rm c}$ is the band's central wavenumber, and $T$ is the half-bandwidth
(\cf Eq.~(\ref{Hj})). 
Let $L_{\cal H}=1/T$ denote the  length scale of the filter's impulse response function.
Apply the filter ${\cal H}$ to the process 
\begin {equation}
\label{LSM_S1_}
\xi (x) =  \sum_{\ell=-\ell_{\rm max}}^{\ell_{\rm max}}  \sigma_\ell(x)  
                          \,\widetilde\alpha_{l} \e^{\i \ell x}
\end {equation}
(we reproduce here Eq.~(\ref{LSM_S1}) for the reader's convenience).
In spectral space, the action of the filter on the signal
$\xi (x) = \sum   \widetilde\xi_n \e^{\i n x}$ amounts to the 
multiplication of the spectral coefficients $\widetilde\xi_n$ by $H(n)$.
The spectral-space representation of $\xi(x)$ is given by Eq.~(\ref{xi_ln}) so 
the filtered process $\varphi = {\cal H}\xi$ reads
\begin {equation}
\label{xi_H}
\varphi (x) = \sum_{n=-\ell_{\rm max}-Q}^{\ell_{\rm max}+Q}  H(n) \,\widetilde\xi_n \e^{\i n x} =
    \sum_{\ell=-\ell_{\rm max}}^{\ell_{\rm max}} \widetilde\alpha_{\ell} \e^{\i \ell x} 
           \sum_{q=-Q}^{Q} H(\ell+q) \,\widetilde\sigma_{\ell q} \e^{\i  q x},
\end {equation}
where, we recall, $\widetilde\sigma_{\ell q}$ are the spectral coefficients 
of the expansion $\sigma_\ell(x) = \sum_{q} \,\widetilde\sigma_{\ell q} \e^{\i  q x}$.
We suppose that  the spatial filter in question is selected such that $L_{\cal H} \ll \Lambda$.
Here $\Lambda$ is the non-stationarity length scale defined on the circle from
\begin {equation}
\label{Lsigma_S1}
\frac{\Ex \xi^2}{\Lambda^2} = \sum_{\ell=-\ell_{\rm max}}^{\ell_{\rm max}}
\Ex \left( \frac{\partial \sigma_\ell}{\partial x} \right)^2
\end {equation}
(\cf Eqs.~(\ref{Lsigma}) and (\ref{Lsigma2})).
Heuristically, $H(\ell+q)$ in Eq.~(\ref{xi_H}) is then
much flatter than $\widetilde\sigma_{\ell q}$ (both being considered as functions of $q$).
As a result,  within the width of  $\widetilde\sigma_{\ell q}$ 
(regarded as a function of $q$ with the fixed $\ell$), it holds that $H(\ell+q) \approx H(\ell)$. 
This suggests the approximation, $\utilde{\varphi} (x)$, 
that results from replacing, in Eq.~(\ref{xi_H}), $H(\ell+q)$ with  $H(\ell)$:
\begin {equation}
\label{LSM_S1_H}
\varphi (x) \approx  \utilde{\varphi} (x) =  \sum_{\ell=-\ell_{\rm max}}^{\ell_{\rm max}}   \sigma_\ell(x)  
                          \,\widetilde\alpha_{l} \, H(l) \e^{\i \ell x}.
\end {equation}
Its spherical counterpart is given in Eq.~(\ref{phi1}).
Now we study the approximation error
\begin {equation}
\label{LSM_S1_H_err}
\delta \varphi(x) = \utilde{\varphi} (x)  - \varphi (x) =  
   \sum_{\ell=-\ell_{\rm max}}^{\ell_{\rm max}}  \widetilde\alpha_{\ell} \e^{\i \ell x} 
    \sum_{q=-Q}^{Q}  \left[ H(\ell) - H(\ell+q) \right] \,\widetilde\sigma_{\ell q} \e^{\i  q x}.
\end {equation}
It has mean zero and variance
\begin {equation}
\label{LSM_S1_H_err2}
\Ex\left[|\delta \varphi(x)|^2 \given \sigma \right] = 
  \sum_{\ell=-\ell_{\rm max}}^{\ell_{\rm max}} \sum_{q=-Q}^{Q}   \sum_{q'=-Q}^{Q} 
     \left[ H(\ell) - H(\ell+q) \right] \left[ H(\ell) - H(\ell+q') \right] 
        \,\widetilde\sigma_{\ell q}   \,\widetilde\sigma^*_{\ell q'} \e^{\i  (q-q') x}.
\end {equation}
We take here expectation over the randomness 
of the  spectral functions  $\sigma_\ell(x)$. 
Since the  vector-valued random process
$\boldsymbol\sigma(x) = \{ \sigma_\ell(x)  \}_{\ell=-\ell_{\rm max}}^{\ell_{\rm max}}$ 
is postulated to be stationary, see  section \ref{sec_LSM_locsta}, it holds that
$\Ex \widetilde\sigma_{\ell q} \widetilde\sigma_{\ell q'}^*=0$
for $q \ne q'$.
As a result, Eq.~(\ref{LSM_S1_H_err2}) simplifies to
\begin {equation}
\label{LSM_S1_H_kappa}
\Ex|\delta \varphi|^2 = \Ex\Ex\left[|\delta \varphi(x)|^2 \,|\, \sigma \right]  =
  \sum_{\ell=-\ell_{\rm max}}^{\ell_{\rm max}} \sum_{q=-Q}^{Q} 
    \left[ H(\ell) - H(\ell+q) \right]^2 \Ex |\widetilde\sigma_{\ell q}|^2.
\end {equation}
Now, we note that for each $\ell$, there exists  $\bar\ell$ such that
$H(\ell+q) - H(\ell) = H'(\bar\ell)q  = \kappa'(\cdot) q/ T \equiv \kappa'(\cdot) q L_{\cal H}$.
Hence
\begin {equation}
\label{LSM_S1_H_kappa2}
\Ex|\delta \varphi|^2  =  L_{\cal H}^2
   \sum_\ell [\kappa'(\cdot)]^2 \sum_q   q^2 \Ex|\widetilde\sigma_{\ell q}|^2 \le
  L_{\cal H}^2 \sum_\ell               \Ex \left( \frac{\partial \sigma_\ell}{\partial x} \right)^2 =
\left( \frac{L_{\cal H}}{\Lambda} \right)^2  \Ex \xi^2,
\end {equation}
where  Eq.~(\ref{Lsigma_S1}) as well as the assumption  $|\kappa'(\cdot)|\le 1$
were used.

Note that 
$\Ex|\utilde\varphi(x)|^2 = \Ex\Ex\left[|\utilde\varphi(x)|^2 \,|\, \sigma \right] 
=\sum H^2(\ell) \Ex|\sigma_\ell|^2$ does not change neither with $\Lambda$ nor with $L_{\cal H}$ 
as long as the mean process spectrum $\Ex\sigma_\ell^2$ is not changed, as we assume. 
Therefore the relative approximation error variance tends to zero as $L_{\cal H} /\Lambda \to 0$ at the
same rate  $(L_{\cal H} /\Lambda)^2$ as the absolute error variance.


Finally, we assess the mean absolute approximation error in Eq.~(\ref{vj}),
$\Ex|\delta^{\rm nsta}(x)|$, where
$\delta^{\rm nsta}(x) = \utilde{v}(x) -v(x)$, 
$v(x)=\Ex(|\varphi(x)|^2 \given \sigma)$, and 
$\utilde{v}(x)=\Ex(|\utilde{\varphi}(x)|^2 \given \sigma)$.
We have 
$|\delta^{\rm nsta}| = \bigl|\Ex\bigl(|\utilde{\varphi}|^2 - |\varphi|^2 \given \,\sigma\bigr) \bigr| \le
 \Ex\bigl( \bigl| |\utilde{\varphi}|^2 - |\varphi|^2  \bigr| \,\given \sigma\bigr) = 
 \Ex\bigl( \bigl| |\utilde{\varphi}| - |\varphi|  \bigr| (|\utilde{\varphi}| + |\varphi|) \,\given \sigma\bigr)
$.
Further, the inequality $\bigl| |a| - |b| \bigr| \le |a-b|$ entails that
$|\delta^{\rm nsta}|  \le
 \Ex\bigl(  |\delta\varphi| (|\utilde{\varphi}| + |\varphi|) \,\given \sigma \bigr)$. 
Next, taking expectation with respect to the randomness of the spectral functions $\sigma$, 
we obtain 
$\Ex|\delta^{\rm nsta}|  \le
 \Ex\bigl(  |\delta\varphi| (|\utilde{\varphi}| + |\varphi|) \bigr)$.
Here, the Cauchy-Schwarz inequality yields
$\Ex|\delta^{\rm nsta}|  \le
 \sqrt{\Ex|\delta\varphi|^2} \sqrt{\Ex \bigl(|\utilde{\varphi}| + |\varphi| \bigr)^2}$.
Finally, assuming that $\Ex \bigl(|\utilde{\varphi}| + |\varphi| \bigr)^2$ is bounded above, we conclude 
from Eq.~(\ref{LSM_S1_H_kappa2}) that
the mean absolute non-stationarity error in $\utilde{v}(x)$, that is, the error in 
the approximate Eq.~(\ref{vj}), is
\begin {equation}
\label{err_v}
\Ex|\delta^{\rm nsta}| = O\left( \frac{L_{\cal H}}{\Lambda}  \right)
\end {equation}
as ${L_{\cal H}}/{\Lambda} \to 0$.

\section{Proof of Proposition \ref{prop_consi}: Consistency of the estimator}
\label{App_consi}

Here, we examine the accuracy of the linear estimator described in section \ref{sec_estm_consi}.
We consider the large sample limit, $K \to\infty$. 
The degree of the spatial non-stationarity is kept fixed in this limit.
We show that the bandpass filters can be selected in such a way that 
the estimation error vanishes, implying consistency of the estimator.

We start by recalling that according to Eqs.~(\ref{vj}) and (\ref{vj2}),
the $j$th band variance can be represented as 
\begin {equation}
\label{vjd}
v_{j} = \utilde{v}_{j} + \delta^{\rm nsta}_j, 
\end {equation}
where
$\utilde{v}_{j}$ is linearly related to the local spectrum and $\delta^{\rm nsta}_j$
is the error of the linear approximation
(here and in the sequel, dependencies on the spatial point $x$ are dropped).
 
Then, following section \ref{sec_estm_consi}, we replace the sum in Eq.~(\ref{vj2}) with the integral:
\begin {equation}
\label{vjI}
\utilde{v}_{j} = \int_{\S^1}  \omega_j(\vartheta; T) \, F(\vartheta) \,\d\vartheta,
\end {equation}
where $\vartheta$ is the log-wavenumber, 
$F(\vartheta)$ represents the local spectrum in the sense that 
$f_\ell =  \sigma_\ell^2 = F(\vartheta(\ell))$,
$\omega_j(\vartheta; T)$ stands for $\frac{2\ell +1}{4\pi}  H_j^2(\ell)$, and 
$T$ is a real number that controls the widths of the functions $\omega_j(\cdot;T)$
(the way it does so is specified below).
The discrepancy $\delta^{\rm nsta}_j$ is studied in \ref{App_accu_est}, where it is generically denoted as 
$\delta^{\rm nsta}$.

We specify the transfer functions $\omega_j(\vartheta,T)$ of the bandpass filters in Eq.~(\ref{vjI})  as follows.
First, we let the number of the filters, $J$, be equal to $M+1$, where
$M$ is the bandwidth of the local spectrum, see Definition \ref{def_smoo}. Second,
we define the function $\Omega: \R \to \R$ as a positive even function that is monotonic on $\R_+$, vanishes at infinity,
and satisfies for some $\beta>0$ 
\begin {equation}
\label{Omega}
\Omega(t) = 1 - |t|^\beta + o(|t|^\beta)
\end {equation}
as $t\to 0$. 
Third, we define 
\begin {equation}
\label{omega}
\omega(\vartheta; T)=  \Omega (\vartheta/T)
\end {equation}
and
let 
\begin {equation}
\label{omj}
\omega_j(\vartheta; T) = \omega (\vartheta_j - \vartheta ;T),
\end {equation}
where $\vartheta_j = \frac{\pi}{M}(j-1)$, $j=1,2,\dots, J$, 
determines the $j$th passband's central log-wavenumber. 
We note that by definition, see Eqs.~(\ref{omega}) and (\ref{omj}), $T$ determines the 
length scale of the resulting filters' impulse response functions:
$L_{{\cal H}_j} = O(T^{-1})$ as $T \to\infty$.
From  Eq.~(\ref{err_v}), we obtain
\begin {equation}
\label{vjd_eps}
 \Ex |\delta^{\rm nsta}_j| = 
O(T^{-1}).
\end {equation}
The data we have are  the sample band variances $d_j$,
which estimate the band variances $v_{j}$ up to the sampling error $\delta^{\rm samp}_j$:
\begin {equation}
\label{vjs}
d_j = v_{j} + \delta^{\rm samp}_j.
\end {equation}
The magnitude of the sampling error can be assessed as
\begin {equation}
\label{sdvs}
\Ex |\delta^{\rm samp}_j|   \le \sqrt{\Ex (\delta^{\rm samp}_j)^2} 
  \sim \sqrt{\frac{2}{K}} \Ex v_j = O(K^{-\nicefrac12})
\end {equation}
as $K \to\infty$.
From Eqs.~(\ref{vjd}), (\ref{vjI}), (\ref{omj}), and (\ref{vjs}), we have
\begin {equation}
\label{Finv}
\int_{\S^1}  \omega(\vartheta_j - \vartheta, T) \; F(\vartheta) \,\d\vartheta = 
d_j - \delta^{\rm samp}_j - \delta^{\rm nsta}_j.
\end {equation}
This is a linear inverse problem to be solved with respect to $F(\cdot)$ from the data $\{d_j\}_{j=1}^J$.
The left-hand side of Eq.~(\ref{Finv}) is the {\em convolution}
$\omega(\cdot; T) * F(\cdot)$  evaluated at $\vartheta_j$.
This convolution can readily be  computed
in Fourier space, see Eq.~(\ref{FFk}), where Eq.~(\ref{Finv}) becomes
\begin {equation}
\label{vFpsi}
\widetilde F_m \, \widetilde{\omega}_m(T)  \propto  
  \widetilde d_m - \widetilde{\delta_m^{\rm samp}} - \widetilde{\delta^{\rm nsta}_m},
\end {equation}
where  the variables with the tilde are Fourier transforms of the respective variables without the tilde
and  $0 \le m \le  M$.
From Eq.~(\ref{vFpsi}), we can write down the solution in Fourier space,
%
$\widetilde{\widehat F}_m  \propto {\widetilde d_m}/{\widetilde{\omega}_m(T)}$,
%
and its error (up to a constant factor):
\begin {equation}
\label{Ferr}
|\delta \widetilde{\widehat F}_m| \le 
\frac{|\widetilde{\delta_m^{\rm samp}|} + |\widetilde{\delta^{\rm nsta}_m}|}{\widetilde{\omega}_m(T)}.
\end {equation}
From Eqs.~(\ref{Omega}) and (\ref{omega}), it follows that
$1/\widetilde{\omega}_m(T) = O(T^{\beta})$ where, we recall, $\beta>0$. 
We substitute this latter asymptotic relation in Eq.~(\ref{Ferr}) and
take expectation of the resulting equation with respect to the distribution of 
the random spectral functions $\sigma_\ell(\cdot) = \sqrt{f_\ell(\cdot)}$ (see section \ref{sec_LSM_locsta}).
We examine the asymptotic behavior of the mean absolute error $\Ex|\delta {\widehat F}(\vartheta)|$ as
$K\to\infty$ and, importantly,  $T=K^\gamma$ (where $\gamma >0$), with
$M=\constant$. 
From  Eq.~(\ref{Ferr}), 
we obtain, again up to a constant factor,
%
%
\begin {equation}
\label{Ferr_j}
\Ex|\delta {\widehat F}(\vartheta)| \le 
\Ex\sum_{m=-M+1}^M |\delta \widetilde{\widehat F}_m| \le 
O(T^{\beta} K^{-\nicefrac12}) + O(T^{\beta-1}). 
\end {equation}
Here the first inequality is because 
$\delta {\widehat F}(\vartheta) = \sum_{m=-M}^M \delta \widetilde{\widehat F}_m \e^{\i m \vartheta}$
and therefore 
$|\delta {\widehat F}(\vartheta)| \le \sum_{m=-M}^M |\delta \widetilde{\widehat F}_m|$.
The second inequality is due to Eqs.~(\ref{Ferr}), (\ref{vjd_eps}),  (\ref{sdvs}), and the assumption that
$M$ is finite and fixed.

From Eq.~(\ref{Ferr_j}), taking $0 < \beta< 1$ and $0 < \gamma <1/(2\beta)$, we have  
$\Ex|\delta {\widehat F}(\vartheta)| \to 0$ as $K\to\infty$.
Recalling that $f_\ell = F(\vartheta(\ell))$, 
we conclude that the local spectrum $f_\ell(x)$ can be estimated from the set of 
sample band variances $d_j(x)$,
and the mean absolute estimation error vanishes as the ensemble size goes to infinity.

It is worth reiterating that this result is obtained in the limit $K \to\infty$ while
the degree of non-stationarity is held fixed.

\section{The Parametric Locally Stationary Convolution Model}
\label{App_pLSM}

Here we introduce a parametric spatial model used in  section \ref{sec_expm_anls}
as a hyperprior model to train the neural  estimator and as a model of  `truth'  
to test the LSEF analysis on the sphere.
The model generates (in a hierarchical manner) realizations of the spatially variable random spectral functions $\sigma_\ell(x)$ and 
realizations of the respective non-stationary random field $\xi(x) \given \sigma$.

\subsection {General design}
\label{App_true_LSM_low_hier}

At the first (\ie the lowest) level in the hierarchy is the random field $\xi(x)$ defined conditionally 
on the spectral functions $\sigma_\ell(x)$ according to Eq.~(\ref{LSM}).

At the second level are the random spectral functions $\sigma_\ell(x)$, which  are
postulated to obey the following parametric model:
\begin {equation}
\label{bnx}
\sigma_\ell(x) = \sqrt{ \frac{c(x)} {1 + \left[\lambda(x) \ell) \right]^{\gamma(x)}}}.
\end {equation}
Here $c(x)$ is the normalizing variable that ensures that 
$\Var\xi(x) =  \sum_\ell \frac{2\ell +1}{4\pi}  \sigma^2_\ell(x) = s^2(x)$ and
$s(x)$, $\lambda(x)$, $\gamma(x)$ are random fields, which we call {\em parameter fields}.

\subsection {Parameter  fields}
\label{App_true_LSM_prm_flds}

The random field
$s(x)$ equals the spatially variable standard deviation of the random field in question. 
$\lambda(x)$ is the local length scale parameter.
$\gamma(x)$ is the local shape parameter.
These three fields are set to be 
non-Gaussian {\em stationary} random fields computed as follows:
\begin {equation}
\label{cnx}
\begin{split}
s(x) = {s}_{\rm add} + {s}_{\rm mult}\cdot g(\ln\varkappa_S\cdot \chi_S(x)),\\
\lambda(x) = \lambda_{\rm add} + {\lambda}_{\rm mult} \cdot g(\ln\varkappa_\lambda\cdot \chi_\lambda(x)),\\
\gamma(x) = \gamma_{\rm add} + \gamma_{\rm mult} \cdot g(\ln\varkappa_\gamma\cdot \chi_\gamma(x)).
\end{split}
\end {equation}
Here $g$ is the nonlinear transformation function defined in the next paragraph and
$\chi_S,\chi_\lambda,\chi_\gamma$ are the three {\em pre-transform}  fields 
(introduced in the next subsection). 
The  pre-transform fields are independent stationary zero-mean Gaussian fields.
The hyperparameters with the subscript  $_{\rm add}$ prohibit unrealistically
small values in the respective parameter fields and, together with 
the hyperparameters with the subscript  $_{\rm mult}$, specify the 
median values of the respective parameter fields.
The hyperparameters $\varkappa_\bullet >0$ affect the magnitude of
deviations of the parameter fields from their median values.

The transformation function $g$ in Eq.~(\ref{cnx}) is taken, following \citet{DSADM},
to be the scaled and shifted  logistic function:
\begin {equation}
\label{logist}
g(z) =   \frac{1+\e^\mathnormal{b}} {1+\e^\mathnormal{b-z}},
\end {equation}
where $b=1$ is the constant. The function $g(z)$ 
behaves like the ordinary exponential function everywhere except for $z \gg b$,
where the exponential growth  is tempered and $g(z) \approx 1+\e^b$.
The reason to replace the more common $\e^{z}$ with $g(z)$ is the desire
to avoid too large values in the parameter fields, which can give rise to unrealistically large spikes in $\xi$.

Due to the nonlinearity of the transformation function $g$, the above
parameter fields $s(x),\lambda(x), \gamma(x)$ are non-Gaussian.
Their point-wise distributions are known as logit-normal or logit-Gaussian.
If some $\varkappa_\bullet=1$, the respective parameter field is constant in space.
The higher  $\varkappa_\bullet$,
the more variable in space becomes the respective parameter.
In the numerical experiments described below, all $\varkappa_\bullet$ 
are the same, equal to $\varkappa$. 
We refer to $\varkappa$ as the 
 {\em non-stationarity strength} hyperparameter because it specifies the degree of spatial  deviations
of the fields $s(x)$, $\lambda(x)$, and $\gamma(x)$ (according to Eq.~(\ref{cnx}))
and thus determines the magnitude of spatial variability in the spectral functions $\sigma_\ell(x)$, see  Eq.~(\ref{bnx}),
which cause non-stationarity.

\subsection  {Pre-transform fields}
\label{App_true_LSM_pre}

The {\em pre-transform} random fields 
$\chi_S(x),\chi_\lambda(x),\chi_\gamma(x)$
are mutually independent zero-mean and unity-variance {\em stationary} Gaussian processes  
whose common spatial spectrum is 
\begin {equation}
\label{f_chi}
f_\ell^\chi \propto \frac{1} {1 + (\Lambda_{\rm NSL} \cdot\ell)^\Gamma},
\end {equation}
where $\Gamma =\gamma_{\rm add} + \gamma_{\rm mult}$,
\begin {equation}
\label{mu_NSL}
\Lambda_{\rm NSL} = L_{\rm med} \cdot  \mu_{\rm NSL}
\end {equation}
is the non-stationarity length scale, 
\begin {equation}
\label{Lmed}
L_{\rm med} = \lambda_{\rm add} + {\lambda}_{\rm mult}
\end {equation}
is the median process length scale,  and $\mu_{\rm NSL}$ is the 
{\em non-stationarity  length} scale  hyperparameter.


Back substitution of the pre-transform random fields 
$\chi_S,\chi_\lambda,\chi_\gamma$ into Eq.~(\ref{cnx}) followed by plugging in the resulting parameter fields 
$s(x)$, $\lambda(x)$, and $\gamma(x)$ into Eq.~(\ref{bnx}) (after
$c(x)$ is adjusted point-wise as explained in  \ref{App_true_LSM_low_hier})
yields the non-stationary random spectral functions $\sigma_\ell(x)$.
Given $\sigma_\ell(\cdot)$, the random field $\xi(\cdot)$ is computed  using Eq.~(\ref{LSM}).

\subsection  {Hyperparameters}
\label{App_true_LSM_hyper}

The default hyperparameters of the  model were
the following:
$\varkappa=2$,  $\mu_{\rm NSL}=3$, ${s}_{\rm add} + {s}_{\rm mult}=1$, 
${s}_{\rm add} = ({s}_{\rm add} + {s}_{\rm mult}) /10$, 
$\lambda_{\rm add} + \lambda_{\rm mult}=\Delta x \cdot 3$,
$\lambda_{\rm add}=\Delta x /3$,  
$\gamma_{\rm add} + \gamma_{\rm mult}=4$, and
$\gamma_{\rm add}=1$.

\subsection  {Local stationarity}
\label{App_true_LSM_locStatio}

The (random) local spectra  $f_\ell$ defined above in this Appendix give rise to a locally stationary 
random field $\xi(x)$ if $\varkappa \approx 1$ 
(low non-stationarity strength)
or  $\mu_{\rm NSL}  \gg 1$ (large non-stationarity length), \cf section  \ref{sec_LSM_locsta}. 

Indeed, if $\varkappa \approx 1$, the parameter fields are nearly constants in space, 
see Eq.~(\ref{cnx}). Therefore, the spatial variability in the local spectrum $f_\ell(x)$ is small, 
implying local stationarity given that  $\mu_{\rm NSL}  \ge 1$.

If $\mu_{\rm NSL} \gg 1$, the  non-stationarity length  $\Lambda_{\rm NSL}$
(Eq.~(\ref{mu_NSL})) is much larger than the 
median local length scale $L_{\rm med}$,
implying local stationarity according to section \ref{sec_LSM_locsta}.

\section{Details on the neural network}
\label{App_NN}

We used PyTorch  \citep{pytorch} in our experiments on the sphere, section \ref{sec_expm_anls},
and R Torch \citep{keydana} on the circle, section \ref{sec_expm_LSEF}.
In the standard feed-forward network, 
the size of the input layer was $J$ (the number of bandpass filters).
The output layer contained  $\ell_{\rm max}+1$ neurons (the size of the local spectrum).
With $J=5$--7 and $\ell_{\rm max}=50$--$60$, we found that the optimal 
number of  hidden fully connected layers was two and the optimal number of
neurons in each of the hidden layers was $120$.

In the hidden layers, the activation function was ReLU \citep{Goodfellow}.
We  tried a few other activation functions (leaky ReLU, tanh, and sigmoid) and found
little difference in the performance of the estimator (not shown).
To make the magnitudes of the inputs less different, we performed the square-root transformation of 
band variances before feeding them to the network (this appeared to be slightly beneficial). 

In the training process, we used the Adam optimizer (a stochastic
gradient-based numerical optimization algorithm  \citep[e.g.][]{Kingma,Goodfellow}.
The network's hyperparameters were taken according to recommendations in \citet{Kingma}.
The size of the minibatch (a random subsample without replacement of the training sample,
used to compute an estimate of the current gradient of the loss function with respect to the network weights)  was tuned to 
be 2500.
The  number of iterations over the whole training sample (epochs) was 30 on the sphere and 200 on the circle.


\bibliography{mybibfile}

\let\thefigureSAVED\thefigure 


\end{document}